\documentclass{article}

\usepackage[final]{neurips_2023}
\usepackage[utf8]{inputenc} % allow utf-8 input
\usepackage[T1]{fontenc}    % use 8-bit T1 fonts
\usepackage{hyperref}       % hyperlinks
\usepackage{url}            % simple URL typesetting
\usepackage{booktabs}       % professional-quality tables
\usepackage{amsfonts}       % blackboard math symbols
\usepackage{nicefrac}       % compact symbols for 1/2, etc.
\usepackage{microtype}      % microtypography
\usepackage{lipsum}
\usepackage{graphicx}
\usepackage{enumitem}
\usepackage{color}
\usepackage{verbatimbox}
\usepackage{listings}
\usepackage{xcolor}
\usepackage[numbers]{natbib}

\definecolor{codegreen}{rgb}{0,0.6,0}
\definecolor{codegray}{rgb}{0.5,0.5,0.5}
\definecolor{codepurple}{rgb}{0.58,0,0.82}
\definecolor{backcolour}{rgb}{0.95,0.95,0.92}

\setlist[itemize]{leftmargin=*}
\setlist[enumerate]{leftmargin=*}
\graphicspath{ {./images/} }

\lstdefinestyle{mystyle}{
  backgroundcolor=\color{backcolour}, commentstyle=\color{codegreen},
  keywordstyle=\color{magenta},
  numberstyle=\tiny\color{codegray},
  stringstyle=\color{codepurple},
  basicstyle=\ttfamily\footnotesize,
  breakatwhitespace=false,         
  breaklines=true,                 
  captionpos=b,                    
  keepspaces=true,                 
  numbers=left,                    
  numbersep=5pt,                  
  showspaces=false,                
  showstringspaces=false,
  showtabs=false,                  
  tabsize=2
}

\usepackage[strict]{changepage}
\usepackage{xcolor}
\usepackage{framed}
\definecolor{demonstrationshade}{rgb}{0.95,0.95,1}
\definecolor{promptshade}{rgb}{0.95,0.95,1}

\definecolor{codegreen}{rgb}{0,0.6,0}
\definecolor{codegray}{rgb}{0.5,0.5,0.5}
\definecolor{codepurple}{rgb}{0.58,0,0.82}
\definecolor{backcolour}{rgb}{0.95,0.95,0.92}

\setlist[itemize]{leftmargin=*}
\setlist[enumerate]{leftmargin=*}
\graphicspath{ {./images/} }

\lstdefinestyle{mystyle}{
  backgroundcolor=\color{backcolour}, commentstyle=\color{codegreen},
  keywordstyle=\color{magenta},
  numberstyle=\tiny\color{codegray},
  stringstyle=\color{codepurple},
  basicstyle=\ttfamily\footnotesize,
  breakatwhitespace=false,         
  breaklines=true,                 
  captionpos=b,                    
  keepspaces=true,                 
  numbers=left,                    
  numbersep=5pt,                  
  showspaces=false,                
  showstringspaces=false,
  showtabs=false,                  
  tabsize=2
}

\makeatletter
\def\thanks#1{\protected@xdef\@thanks{\@thanks
        \protect\footnotetext{#1}}}
\makeatother

\title{BattleAgent: Multi-modal Dynamic Emulation on Historical Battles to Complement Historical Analysis}

\author{%
 Shuhang Lin$^*$ \\
 Rutgers University
 \And
 Wenyue Hua$^*$ \\
 Rutgers University
 \And
 Lingyao Li\\
 University of Michigan
 \And
 Che-Jui Chang\\
 Rutgers University
 \And
 \hspace{-20pt}Lizhou Fan\\
 \hspace{-20pt}University of Michigan
 \And
 \hspace{-5pt}Jianchao Ji \\
 \hspace{-5pt}Rutgers University
 \And
 Hang Hua \\
 University of Rochester
 \And
 Mingyu Jin \\
 Rutgers University
 \And
 \hspace{-5pt}Jiebo Luo \\
 \hspace{-5pt}University of Rochester
 \And
 \hspace{-5pt}Yongfeng Zhang\thanks{$^*$Shuhang Lin and Wenyue Hua have equal contributions. \textbf{Author Emails}: \{shuhang.lin,wenyue.hua, yongfeng.zhang\}@rutgers.edu} \\
 \hspace{-5pt}Rutgers University
}

\begin{document}

\maketitle

\begin{abstract}
This paper presents \textbf{BattleAgent}, a detailed emulation demonstration system that combines the Large Vision-Language Model (VLM) and Multi-Agent System (MAS). This novel system aims to simulate complex dynamic interactions among multiple agents, as well as between agents and their environments, over a period of time. It emulates both the decision-making processes of leaders and the viewpoints of ordinary participants, such as soldiers. The emulation showcases the current capabilities of agents, featuring fine-grained multi-modal interactions between agents and landscapes. It develops customizable agent structures to meet specific situational requirements, for example, a variety of battle-related activities like scouting and trench digging. These components collaborate to recreate historical events in a lively and comprehensive manner while offering insights into the thoughts and feelings of individuals from diverse viewpoints. The technological foundations of BattleAgent establish detailed and immersive settings for historical battles, enabling individual agents to partake in, observe, and dynamically respond to evolving battle scenarios. This methodology holds the potential to substantially deepen our understanding of historical events, particularly through individual accounts. Such initiatives can also aid historical research, as conventional historical narratives often lack documentation and prioritize the perspectives of decision-makers, thereby overlooking the experiences of ordinary individuals. This biased documentation results in a considerable gap in our historical understanding, as many stories remain untold. BattleAgent leverages the current advancements in Artificial Intelligence (AI) to provide some insights to bridge this gap. It illustrates AI's potential to revitalize the human aspect in crucial social events, thereby fostering a more nuanced collective understanding and driving the progressive development of human society. Quantitative evaluations are computed on the final emulation result, showing reasonable performance and effectiveness of the approach. The data and code for this project are accessible at \url{https://github.com/agiresearch/battleagent} and the demo will be released in one  month.
\end{abstract}

\section{Introduction}
An agent is defined as a system that has the ability to perceive its surroundings and make informed decisions based on these perceptions to accomplish specific objectives \cite{xi2023rise}. The recent progress in Large Language Models (LLMs) \cite{zhao2023survey, fan2023bibliometric} has demonstrated impressive reasoning capabilities \cite{huang2022towards, jin2024impact}, indicating their potential to serve as the foundation for agents. These models have shown remarkable proficiency in following instructions \cite{zhou2023instruction, zeng2023evaluating}, interpreting commands, and emulating human reasoning and learning processes \cite{wang2024llms, wang2024tool, shen2024small}. Additionally, the development of large Vision Language Models (VLM) \cite{zhang2024vision} has facilitated the creation of various agent applications that support multi-modal information interaction \cite{durante2024agent, xie2024large}. When combined with external tools, either physical or virtual, these agents employ LLMs or VLM as their reasoning backbone to determine how tasks should be addressed, how tools should be utilized, and what information should be retained in memory. This enhancement equips agents to manage an array of natural language processing tasks and engage with their environment using language.

A multitude of agent applications have been created using LLM and VLM, with a focus on enhancing reasoning \cite{du2023improving, chan2023chateval, sun2023corex, liang2023encouraging}, production capabilities \cite{hong2023metagpt, liu2023bolaa, ge2023openagi, yang2023appagent, mei2024llm, ge2023llm}, gaming \cite{gong2023mindagent, xu2023exploring, lan2023llm, hu2024survey}, and social simulation \cite{pang2024self, zhou2024real, sreedhar2024simulating, xie2024can, hua2023war}, among others. WarAgent \cite{hua2023war} is the pioneering LLM-based MAS simulation of historical events, examining the behaviors of systems at the macro level, such as nations and governments, rather than the micro-level simulation of detailed and dynamic events occurring during battles or individual experiences in such dynamic time periods. Therefore, BattleAgent, building on the foundation laid by WarAgent in historical event simulation, investigates the potential of LLM and VLM for detailed historical situation recovery and the exploration of individual experiences within the simulation.

\begin{figure}
    \centering
    \includegraphics[scale=0.5]{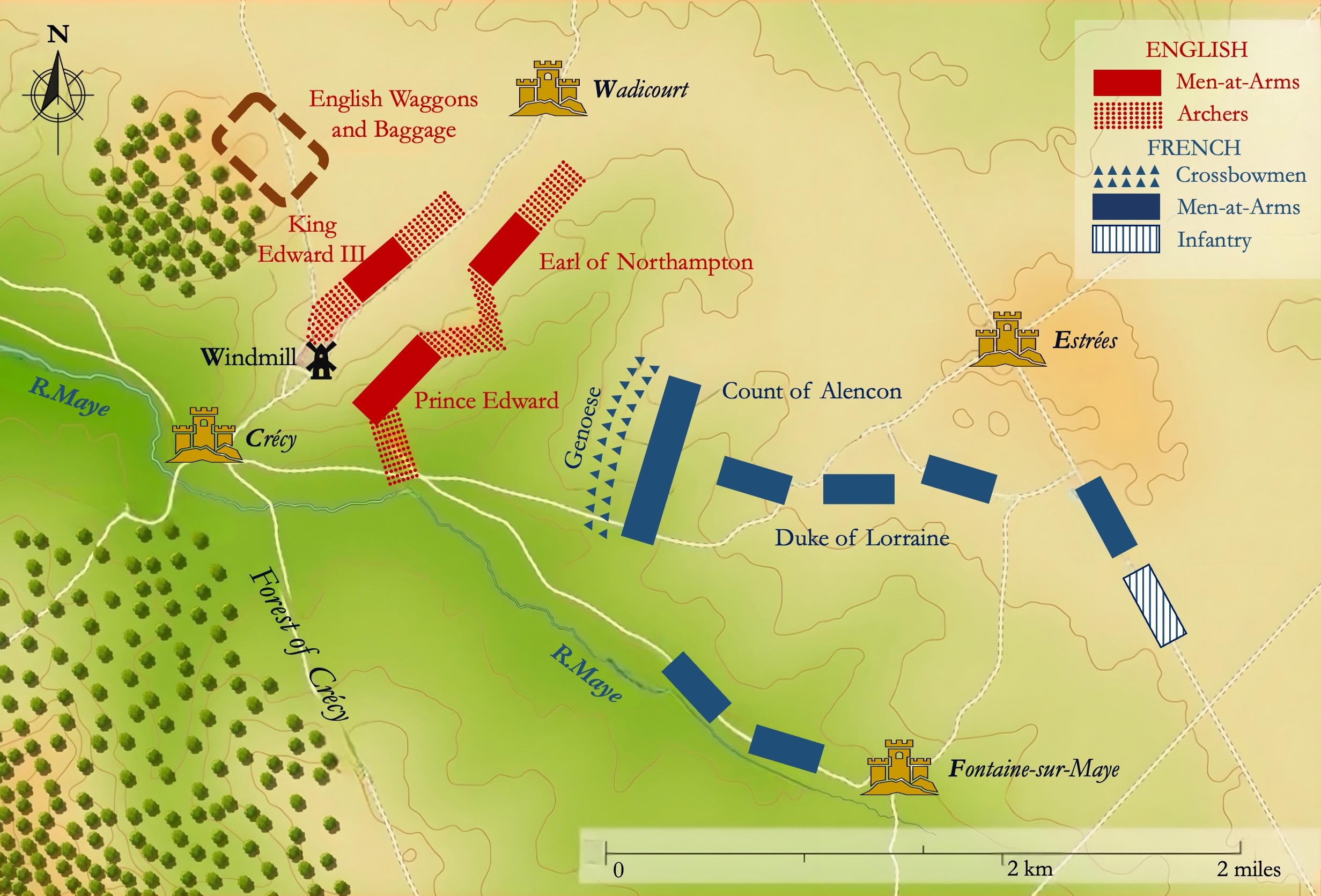}
    \caption{Demonstration of the emulated Battle of Crécy, 1346: Troop formations and movements depicting the positions of the English and French forces during the historical engagement, with key locations and leaders marked.}
    \small\textsuperscript{Image adjusted from 
    \url{https://the-past.com/feature/the-battle-of-crecy-26-august-1346/}}
    \label{fig:main_demonstration}
\end{figure}

The study of history has long been a pursuit to understand the human experience through the lens of past events. Traditional historical narratives often focus on the perspectives of leaders and decision-makers, leaving the experiences of ordinary individuals in the shadows. This selective approach to history has created a significant gap in our understanding, as the stories and experiences of the common participants, such as soldiers, are frequently overlooked. The motivation behind this research is to address this imbalance and provide a more comprehensive view of historical events by leveraging advancements in AI. Oral history \cite{shopes2011oral, portelli2002makes} has been one method used to capture the experiences of individuals, offering a more personal account of historical events. However, this approach is limited to recent history and is constrained by the availability of witnesses, often leaving many details undiscovered. As we move further away from events in time, the voices of those who lived through them fade, and with them, the richness of history's tapestry. 

In response to these challenges, our study introduces BattleAgent, a novel emulation framework that utilizes LMM-based MAS for the detailed reconstruction of historical events, with an emphasis on delineating the experiences of ordinary individuals, notably soldiers.  BattleAgent emulates historical combats within complex terrains and hierarchical command structures, incorporating sophisticated military logistics and strategic planning. Central to our model, we meticulously craft 30 individual soldier agents, each endowed with a richly detailed background and a distinct personality, thereby infusing them with vibrancy and depth. BattleAgent is designed to emulate and document the experiences of these agents, capturing their actions, injuries, emotional responses, and psychological states throughout the course of the battles. By analyzing these elements, we generate personalized narratives that reflect the multifaceted experiences of ordinary individuals engaged in warfare. This approach not only immortalizes the actions and sentiments of these agents but also furnishes a nuanced, individualized perspective on the common soldier's experience within the broader tapestry of historical conflict.

To emulate such a complex scenario, our emulation incorporates the following three key features:
\begin{itemize}
    \item \textbf{Enhanced 2-D Realism Features}: BattleAgent simulates detailed interactions within environments, including terrain engagement, temporal progression, and interactions between agents.
    \item \textbf{Immersive Multi-agent Interactions}: It integrates MAS to facilitate dynamic interactions among agents in battle emulations, accurately reflecting the historical milieu and the intricacies of military engagements, from strategic maneuvers to logistical considerations and communication dynamics.
    \item \textbf{Dynamic Agent Structure}: The framework introduces adaptable agent configurations and multi-modal interactions. The system can ``self improvise'' its structure to fork, merge, and prune agents to continuously maintain the emulation effectiveness. It boasts the capability to autonomously adjust its architecture to optimize emulation fidelity.
\end{itemize}

The contributions of our study to historical analysis and society can be summarized as follows:
\begin{itemize}
    \item \textbf{Emphasis on individual perspectives and granularity}: Providing a platform for the voices of common people to be heard and understood in historical events. This platform aims to enhance the accuracy of historical reconstructions by incorporating individual perspectives. 
    \item \textbf{Connection and resonance with the past}: Helping to prevent future conflicts by learning from the detailed analysis of past mistakes and human costs. This platform fosters empathy and a deeper connection to the past by humanizing the experiences of those involved in historical battles. 
    \item \textbf{Educational tool for understanding history}: Providing an educational tool to help people understand the intricacies of history and the harsh realities of historical events. Its immersive and interactive platform can foster empathy and a more nuanced perspective on the past, making it a valuable resource for students and history enthusiasts. 
    \item \textbf{Potential as a next-generation game engine}: Providing a fully automated process to create immersive and dynamic historical emulations, making it a potential next-generation game engine. By using LLM-based agents and VLM-based agents, it can generate detailed and realistic environments, characters, and events, offering a unique and engaging gaming experience.
\end{itemize}

\section{Related Work}

\subsection{Multi-Agent System}

MAS has revolutionized the landscape of AI, offering a platform for simulating complex interactions and scenarios \cite{hua2023war}. With the development of the reasoning intelligence of LLMs, especially their outstanding reasoning ability in complex scenarios \cite{fan2023nphardeval, fan2024nphardeval4v, zhang2024llm}, the integration of MAS into AI systems has demonstrated their versatility and efficacy. The initial classification of MAS into reasoning-enhancement, non-player character (NPC) multi-agent players, and production-enhancement systems has been foundational in understanding their diverse applications. Notable developments such as LLM-Debate \cite{du2023improving}, ChatEval \cite{chan2023chateval}, and MAD \cite{liang2023encouraging} have significantly advanced reasoning-enhancement systems. Similarly, in NPC multi-agent systems, the emergence of Generative Agents \cite{park2023generative} and GPT-Bargaining \cite{fu2023improving} has paved the way for more human-like agent behaviors. In the production-enhancement domain, innovations like MetaGPT \cite{hong2023metagpt} and OpenAGI \cite{ge2023openagi} have streamlined and enhanced collaborative efforts in software development. Many works have also explored agents' potential in scientific experiments \cite{bran2023augmenting, hua2024trustagent, chen2023towards} setting.

In the context of humanities and historical research, the WarAgent \cite{hua2023war} initiative has exemplified the application of LLM-based MAS for simulating international conflicts, with each agent representing a different country to explore the dynamics of international relations and conflicts. Building on the humanitarian insights gleaned from WarAgent, our study seeks to refine this approach by concentrating on the granular emulation of historical scenarios from the vantage point of ordinary individuals. This shift towards focusing on the micro-level experiences within historical events aims to provide a more detailed and empathetic understanding of the past, leveraging the advancements in MAS and LLM technologies to capture the nuanced perspectives of common people in historical narratives. By doing so, we aspire to enrich the tapestry of historical understanding with a deeper, more inclusive examination of human experiences during pivotal moments in history.

Recent advancements in multi-modal multi-agent AI systems have further expanded the capabilities of MAS. AppAgent \cite{AppAgent} demonstrates the use of multimodal agents as smartphone users, enhancing our understanding of human-computer interactions. The integration of generative AI and multi-modal agents in AWS \cite{AWS} has unlocked new potentials in financial markets. LLaVA-Plus's contribution \cite{LLaVAPlus} in teaching agents to use various tools has opened up new avenues for agent adaptability and functionality. Furthermore, the implementation of ``multimodal chain-of-action agents'' \cite{CoA} has provided a novel perspective on agent interaction with digital interfaces, contributing to more intuitive user interface designs and realistic simulations in digital realms.

BattleAgent emulation is the first large multimodal model-based multi-agent application that introduces a novel quantitative dimension to historical and humanities studies and underscores the broader impact of AI in understanding human history and shaping future scenarios. By exploring alternative historical pathways and key determinants, our work demonstrates the significant contributions of LMM and MAS in enhancing our comprehension of the past and potentially guiding a more informed and peaceful future.

\subsection{Challenges in Granular Historical Analysis}

The pursuit of simulating historical events using computational methods has evolved significantly over the years. Beginning with human simulations, transitioning to human-program hybrid systems, and finally arriving at fully computerized simulations, each stage has brought unique insights and challenges \cite{hua2023war}. Human simulations, as outlined by Dickson \cite{dickson2002road}, provided a foundational approach. In educational scenarios, such simulations involved role-playing exercises, enabling students to delve into the complexities of historical events, such as the United States' entry into WWI. The advent of human-program hybrid systems, exemplified by the Inter-Nation Simulation model \cite{guetzkow1963simulation} and its various applications \cite{hermann1967attempt}. These systems merged human decision-making with computational processes, creating a more dynamic and interactive environment for simulating international conflicts. However, the reliance on human input still presented limitations in terms of scalability and the possible depth of analysis. 

In the past decade, there has been a significant shift in leveraging computing power to create more sophisticated simulations. The OneSAF Objective System (OOS) \cite{tollefson2008onesaf} and the JAVA-based simulation of the Bay of Biscay submarine war \cite{hill2004using} are prime examples. These simulations used detailed models of military operations and game theory, enhancing the accuracy and depth of historical analysis. More recently, through the development of generative AI methods, high-level simulation of social system dynamics becomes a reality \cite{ghaffarzadegan2023generative}. Despite these advancements, the complexity of human behavior and the vastness of historical data remained challenging to fully encapsulate in these models. Moreover, while fully computerized simulations can achieve the most detailed and accurate simulation among all of the three stages, they still are focuses on the panoramic and high-level simulation of historical analysis, often unable to delve deeper into the individual witnesses' reflection and granular analysis beyond countries' or famous leaders' perspectives.

The ideas of ``agent-based computational models'' and ``generative social science'' have been well-known theories before the start of the 21st century \cite{e21pstein1999agent}. Rule-based agents, while can reconstruct complex social behaviors, to some extent, to investigate ``backward to the future'' \cite{ghaffarzadegan2023generative}, are often unable to maintain human-like intelligence, which is the key to simulating and understanding human society. Granular historical analysis with generative methods seems to be unrealistic even with the most cutting-edge Computational Social Science (CSS) methodologies. The current landscape of CSS, particularly in fields like sentiment analysis \cite{fan2020stigmatization, yin2020using, li2023chatgpt}, primarily operates on contemporary data sources. This presents a significant challenge in historical analysis, as historical data often lacks the granularity and digital format required for computational analysis. 

Our research confronts this challenge by employing an LLM within a MAS framework. This approach represents a novel step in historical simulation, blending the comprehensive data processing capabilities of modern AI with the intricate modeling of MAS. This integration marks a significant departure from traditional methods, as it attempts to overcome the limitations of data scarcity and quality in historical research. By utilizing advanced language models, we can infer, reconstruct, and simulate historical narratives and events with a level of depth and accuracy previously unattainable. Thus, we refer to this granular simulation approach as ``history emulation''. 

Our MAS framework not only models individual agents and their interactions but also incorporates broader socio-political and economic contexts derived from limited historical data. This approach allows for a more nuanced and granular exploration of historical events, shedding light on the complex interplay of factors that shaped these occurrences. Our work, therefore, stands at the forefront of historical emulation, or as we redefined ``history emulation,'' offering a unique blend of AI-powered analysis and traditional historical scholarship. This synergy aims to provide new perspectives on historical events, facilitating a granular, diverse, and deeper understanding of the past and its implications for the future.

\section{Emulation Setting}
This section outlines the emulation framework and setting for our research demonstration. We commence with an exposition of the historical context of the four significant European battles that our emulation seeks to emulate: the Battle of Crécy, the Battle of Agincourt, the Battle of Poitiers, and the Battle of Falkirk. Each battle has been selected for its notable use of cold weapons and the strategic bipartite confrontations that characterized warfare during their respective periods. Building upon the historical context, we elaborate on the configuration of agents and their designated roles within our emulation framework. Our model incorporates two distinct categories of agents to capture the complexity of the battlefield: commanding agents and soldier agents. This dual-layered agent structure enhances the emulation's fidelity, offering nuanced insights into the command and control hierarchies, as well as the individual soldier experiences of historical warfare. To accurately emulate the historical engagements, each agent type has a specific set of actions available.

\subsection{Battle Histories}
\paragraph{Battle of Crécy} \cite{burne2016crecy} a pivotal clash during the Hundred Years' War, occurred on 26 August 1346 in northern France. The English forces under King Edward III confronted the French army led by King Philip VI. The English army is believed to have consisted of  \textit{around 10,000 to 15,000} men, while the French forces are estimated to have been  \textit{between 20,000 and 35,000} strong. As the English army marched through northern France, they were assaulted by the French, leading to a decisive English victory marked by substantial French casualties. As for casualties, the English losses were relatively light, with estimates ranging from \textit{a few hundred to around 2,000} men. On the other hand, the French suffered heavy losses, with estimates suggesting that  \textit{between 10,000 and 15,000} French soldiers were killed, including many high-ranking nobles. The Battle of Crécy marked a significant turning point in the Hundred Years' War and demonstrated the effectiveness of the longbow against traditional knightly charges.

\begin{figure}[!ht]
    \centering
    \includegraphics[scale=0.15]{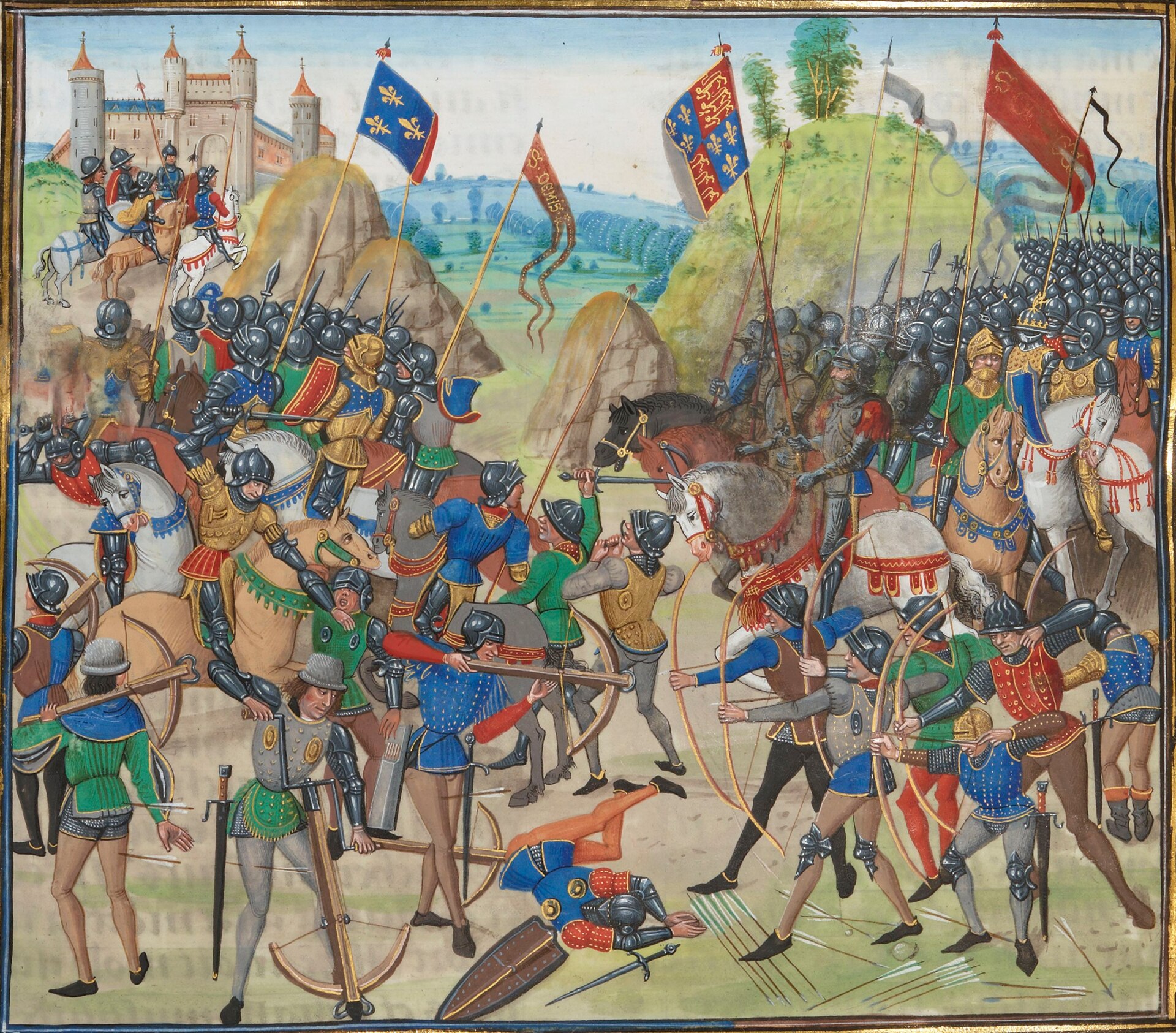}
    \caption{Battle of Crécy, from an illuminated manuscript of Jean Froissart's Chronicles.}
    \label{fig:crecy}
\end{figure}

\paragraph{Battle of Agincourt} \cite{curry2000battle} fought on 25 October 1415 near Azincourt in northern France, stands as a landmark English triumph in the Hundred Years' War. The English army consisted of \textit{around 6,000 men}, primarily made up of archers armed with longbows. The French, on the other hand, had a force of \textit{approximately 12,000 to 36,000} men, composed of knights, men-at-arms, and infantry. Estimates of casualties vary, but it is generally agreed that the English suffered relatively light losses, with \textit{up to 600 men killed}. In contrast, the French suffered heavy casualties, with estimates suggesting that \textit{between 4,000 and 10,000} French soldiers were killed, including many high-ranking nobles. The confrontation took place on Saint Crispin's Day and ended with a surprising victory for the outnumbered English forces. This victory significantly elevated English morale and status, inflicted severe damage on France, and initiated a phase of English ascendancy in the war that persisted for 14 years. This period ended with England's defeat at the hands of France in 1429 during the Siege of Orléans.

\paragraph{Battle of Poitiers} \cite{given2001edward} On 19 September 1356, the Battle of Poitiers was waged between the French army, commanded by King John II, and an Anglo-Gascon contingent led by Edward, the Black Prince. The battle unfolded in western France, near Poitiers, where a French force numbering \textit{between 14,000 to 16,000} men launched an attack against a strategically fortified position held by a \textit{6,000}-strong Anglo-Gascon army. It is generally agreed that the English suffered relatively light losses, with \textit{around 40} men-at-arms killed. In contrast, the French suffered heavy casualties, with estimates suggesting that \textit{around 4,500} French soldiers were killed, including many high-ranking nobles.

\paragraph{Battle of Falkirk} \cite{kiffer2019battle} a major engagement in the First War of Scottish Independence, took place on 22 July 1298. The English army, commanded by King Edward I, achieved a significant victory over the Scots, who were led by William Wallace. The English army consisted of \textit{around 12,000 to 15,000 men}, primarily made up of heavy infantry and archers. The Scottish forces were believed to have numbered \textit{around 5,000 to 8,000 men}, composed mainly of spearmen and some cavalry. English had around \textit{2,000} men killed. In contrast, the Scottish suffered heavy casualties, with estimates suggesting that between \textit{2,000} Scottish soldiers were killed, including many high-ranking nobles. The defeat at Falkirk led to Wallace's resignation as Guardian of Scotland.

\subsection{Agent Definition}
Our basic emulation setting contains the profile definition of agents and their action space. Our emulation encompasses two types of agents:
\begin{enumerate}
    \item Commanding Agents: These agents symbolize the individuals or collective entities responsible for strategic decision-making in the theater of war. They are programmed to emulate the tactical acumen and leadership styles of historical commanders, thereby influencing the broader outcome of the simulated conflict. 
    \item Soldier Agents: These agents personify the rank-and-file soldiers who executed the battle plans on the ground. Each soldier agent is equipped with a comprehensive profile that includes personal history, psychological traits, and potential responses to combat stimuli.
\end{enumerate}

\paragraph{Commanding Agents} These agent profiles include general information about an army (a group of soldiers). Decisions and Strategies of the commanding agent will be made based on the whole army information, which includes the following aspects:
\begin{enumerate}
    \item ID: The ID of a commanding agent is represented by a hash code that is generated to uniquely identify each commanding agent within the emulation sandbox. This is necessary due to the dynamic agent structure employed in our emulation, which allows for the creation of additional agents beyond the initial (two) commanding agents as the emulation progresses. The use of a hash code ensures that each agent can be accurately identified and tracked throughout the course of the emulation.
    \item Military Command Structure: This involves the hierarchical organization and leadership dynamics within each military faction.
    \item Morale and Discipline: An assessment of the troops' psychological readiness, their discipline levels, and overall morale. 
    \item Military Strategy: The overarching tactical approaches and plans employed by each side in the conflict.
    \item Military Capability: An inventory of the weapons and defense tools at each side's disposal.
    \item Force size and composition: This aspect includes the total number of soldiers and their composition including information about the types of troops, their roles, and their proportions in the overall force. 
    \item Location: The current location of the agent is represented by its coordinates. These coordinates provide a precise indication of the agent's position within the simulated environment, allowing for accurate tracking and analysis of its movements and interactions with other agents and the environment.
\end{enumerate}

Below is an example of an English army agent which is initialized at the very start of the emulation for the Battle of Crécy:
\begin{lstlisting}[language=HTML, caption=An example of commanding agent profile]
1. ID: ARMY-13e370a8

2. Command Structure: experienced commanders with significant autonomy

3. Morale and Discipline:
(1) High morale and strict discipline
(2) Enhanced by tactical innovations and effective use of the longbow

4. Military Strategy:         
(1) Aggressively defensive posture, emphasizing strategic high ground for offensive strikes. Vigorously exploits terrain advantages for combative engagements.         
(2) Forceful application of longbows, enabling a confrontational yet adaptive offense, underscored by ambitious tactical innovations.

5. Military Capability:
(1)Longbow: Exhibits a rapid rate of firing and extensive range, with the capability to pierce armor.
(2)Gunpowder Weapons: Encompasses a range of types, incorporating dismounted men-at-arms and selective deployment of cannons.

6. Force Size and Composition:
Size: 10,000. Includes men-at-arms, longbowmen, hobelars, and spearmen.

6. Armament and Protection:
(1) Armor: Men-at-arms wore quilted gambeson under mail armor, supplemented by plate armor, bascinets with movable visors, and mail for throat, neck, and shoulders.
(2) Shields: Heater shields made from thin wood overlaid with leather.
(3) Weapons: included lances used as pikes, swords, and battle axes.
(4) Special Weapon: Longbow.

7. Location: [0, 0]
\end{lstlisting}

\paragraph{Soldier Agents} Soldier agents are characterized by a wide range of attributes, including name, age, familial ties, occupation, personality traits, social standing, potential health conditions, physical fitness, hobbies and interests, conversational style, unique idiosyncrasies, and personal secrets or controversies. These attributes are designed to provide a comprehensive and nuanced representation of the soldier agents.

\subsection{Anonymization of Battle Details}
To prevent providing explicit hints to the LLM and VLM regarding the specific battle being emulated, in case it has memorized certain battle information, we anonymize various battle details. This includes the names of countries, leaders, specific year and date of the battle, and location names. By doing so, we ensure that the LLM's decisions are based solely on the information provided in the prompt and not influenced by any prior knowledge it may have. 

\subsection{Action Space} 
\paragraph{Action Space of Commanding Agents} 
Our emulation framework is characterized by its comprehensive detail, featuring an action space that encompasses 51 distinct actions. Agents within the emulation have the flexibility to select any combination of these actions at each decision point. The actions available in the action space are organized into seven categorically distinct groups:
\begin{enumerate}
    \item Reposition. This category includes actions that involve the movement of an army or a subsection thereof to a different location: \textit{Reposition Forces, Create Decoy Units}
    \item Preparation. Actions in this group are geared towards readying forces for an impending attack: \textit{Deploy Longbows, Rally Troops, Employ Artillery, Use of Gunpowder Weapons, Resupply Archers, Destroy Enemy Morale, Deploy Archers in Flanking Positions, Organize Night Raids, Organize Raiding Parties, Digging trenches}
    \item Attack. This group encapsulates a variety of common attack strategies, such as skirmishing, ambushing, besieging, cavalry charges, and direct firing, among others: \textit{Initiate Skirmish, Charge Cavalry, Ambush Enemy, Launch Full Assault, Archery Duel, Siege Tactics, Hand-to-Hand Combat, Counterattack, Conduct Reconnaissance, Direct Artillery Fire, Engage in Siege Warfare, Execute Flanking Maneuvers, Use Cavalry for Shock Tactics, Employ Archers Strategically}
    \item Defense. Encompasses actions such as shielding, fortification, and the creation of obstacles: \textit{Construct Defenses, Prepare Defenses, Develop Counter-Siege Measures, Form Defensive Shields, Establish Defensive Fortifications, Fortify Rear Guards, Fortify Position, Create Obstacles for Enemy Cavalry, Form Defensive Pike Formations, Set Traps}
    \item Observation. Focused on gathering information about the surrounding area and the current situation of the enemy: \textit{Scout Enemy Position, Gather Intelligence, Intercept Enemy Supplies, Establish Communication Lines}
    \item Retreat. Actions related to strategic withdrawal in the face of adverse conditions: \textit{Retreat and Regroup, Tactical Retreat, Plan Feigned Retreat}
\end{enumerate}

Each action requires 3 inputs: initiator, location, and recipient. For each action, the agent is required to (1) identify the specific agent or entity responsible for emulation of the action. It defines who is carrying out the action within the emulation. (2) the geographical point or area within the emulation where the action takes place. It is crucial for contextualizing the action within the broader landscape of the battle scenario. (3) the target or beneficiary of the action. It specifies towards whom or what the action is directed, whether it is an opposing force, a specific unit, or another entity within the emulation.

\paragraph{Action Space of Soldier Agents}
These soldier agents follow the orders given by their commanding agents and are affected by combat outcomes based on the general fatality rate, which is calculated according to the actions taken, the size of the armies, and the types of weapons involved. By going through this process, the individual agents provide a more granular view of the battlefield dynamics, allowing for a better understanding of the experiences and feelings of soldiers on the ground.

\section{Emulation Sandbox}
In our emulation framework, we concentrate on a relatively straightforward scenario: a bipartite battle. The process begins with the establishment of the geographical context for the entire scenario, both textual description as well as a visual map. Subsequently, we define the two initial opposing commanding agents, considering seven profile aspects to ensure a comprehensive and realistic representation. Notice that in this section, we simply use ``agent'' to refer to ``commanding agent''.

\subsection{Time and Space in Sandbox}
In order to accurately simulate the dynamics of historical battles, it is crucial to effectively manage the time and space within the sandbox environment. In this section, we will discuss our approach to time and space management in the sandbox.

\paragraph{Quantized Time Management}
The battlefield environment is characterized by continuous dynamic changes as shown in both Figure \ref{fig:general_sandbox} and Figure \ref{fig:battle_field}, with agents frequently altering their actions and positions. To accurately emulate these dynamics while preserving the discrete decision-making process in our agent-based emulation, we employ a time quantization approach. Specifically, we discretize the continuous flow of time \cite{matsuoka2001computer, al2011effects} into 15-minute intervals. For each quantized time block, agents have the flexibility to either maintain their current action or adapt their actions based on the unfolding situation.

\paragraph{Coordinate Generation based on Map}
We obtain the initial map of the battlefield from historical documents \footnote{\url{https://en.wikipedia.org/wiki/Battle_of_Crecy}, \url{https://en.wikipedia.org/wiki/Battle_of_Agincourt}, \url{https://en.wikipedia.org/wiki/Battle_of_Falkirk}, \url{https://en.wikipedia.org/wiki/Battle_of_Poitier}}. To generate the coordinates from the original image, we use one army position as the reference point, designated as the (0,0) position. We then use a scale of 10 yards as one unit of the coordinate system. The coordinates of key landscapes on the map are estimated and expressed with a range in the textual description.

For example, in the Battle of Crécy, the coordinates of a river are generated as follows:

\begin{lstlisting}[language=HTML, caption=An example of coordinates of a river in the battlefield of Battle of Crécy]
RiverB: {
path: 
    {
    start: [-100, -200],
    end: [100, 50],
    description: Flow path of the river begins from the southwest corner towards the northeast.
    },
properties: 
    {
    type: River
    },
description: Meandering waterway providing natural boundaries and obstacles.
}
\end{lstlisting}

The coordinates of these key positions, including both key landscapes and existing agents on the battlefield, serve as reference points for agent movement decisions. When making decisions on agent movement, the agent will refer to the coordinates of these key positions and choose the target position to move to. This approach enables agents to navigate the battlefield in a realistic and contextually appropriate manner, taking into account the presence of natural boundaries and obstacles.

\subsection{General Sandbox Process}
\begin{figure}[!ht]
    \centering
    \includegraphics[scale=0.5]{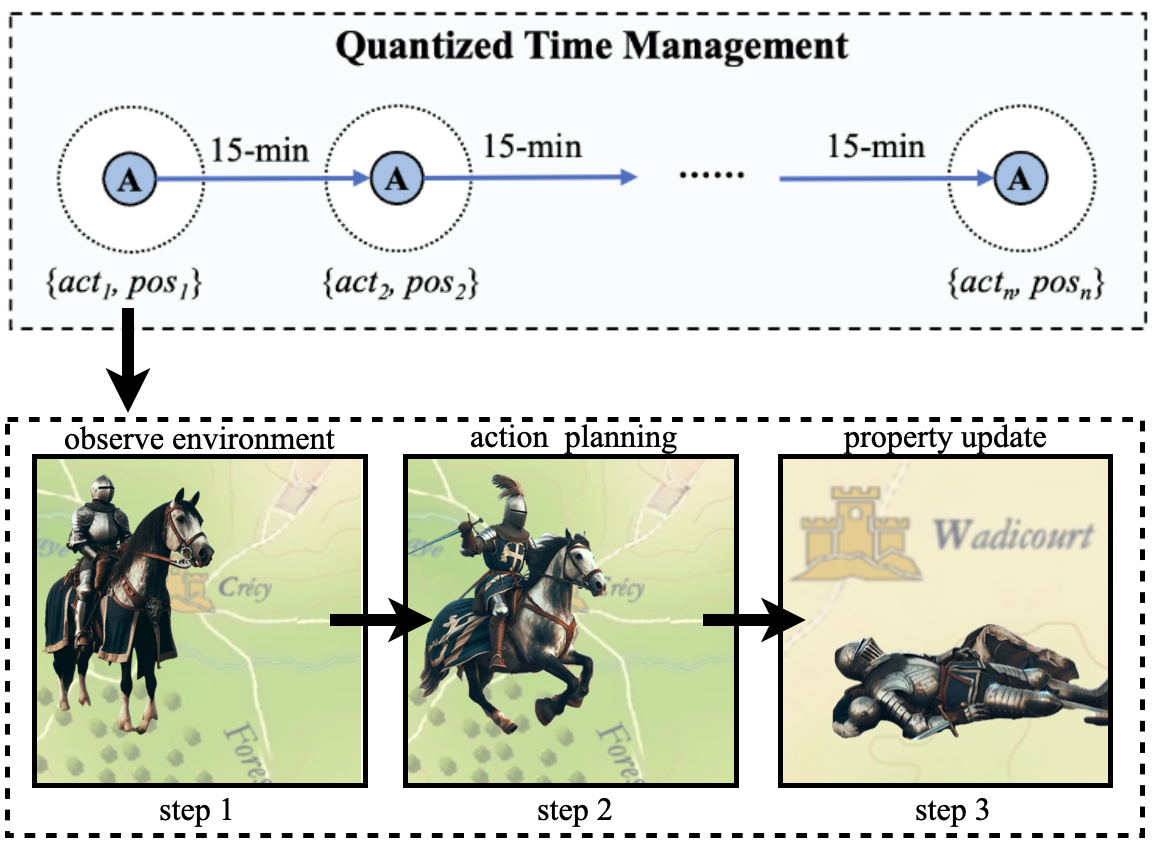}
    \caption{General Sandbox Process: The agent actions and properties are updated every 15 minutes, and for each 15-minute interval, each agent observes the environment, plans its actions, and acts based on its updated properties. This process is repeated for the duration of the simulation to emulate the dynamics of historical battles.}
    \label{fig:general_sandbox}
\end{figure}

Here we provide a very simple and crude overview of the emulation sandbox, as presented in Figure \ref{fig:general_sandbox}. We initiate the emulation based on historical map which contain information about geography as well as the position of the armies. The following represents a high-level overview of the steps involved in the emulation process:
\begin{itemize}
    \item Step 1: Each agent starts by observing its surroundings and gathering information. This observation process involves text-based reasoning from the agent's prompt as well as obtaining visual information from the map.
    \item Step 2: Based on the gathered information, each agent decides on its actions, such as preparing for battle (e.g., digging trenches, reinforcing troops), collecting further information, or making organizational agent structure changes like forking, merging, or being pruned to dynamically create new agents or diminish existent agents.
    \item Step 3: For every 15-minute interval in the emulation sandbox, the general information, including agent locations and properties, is updated according to the actions taken by all agents.
    \item The process then loops back to Step 1, with agents continuing to observe, make decisions, and act based on the updated information and evolving battlefield situation.
\end{itemize}

By following this iterative process, our agent-based emulation can effectively capture the complex dynamics of the battlefield while maintaining the discrete decision-making process inherent to the emulation.

\subsection{Detailed Emulation Process}

In this subsequent section, we delve into a more comprehensive explication of the general sandbox process previously introduced, elaborating on several key aspects. These include the management of time in a quantized manner, the process of making observations grounded in the map, the planning and execution of actions by the agents, the evaluation of casualty for each agent at every quantized time interval and the methodology behind updating the properties of each agent.

\subsubsection{Observation based on Map Information}

\begin{figure}[!ht]
    \centering
    \includegraphics[scale=0.6]{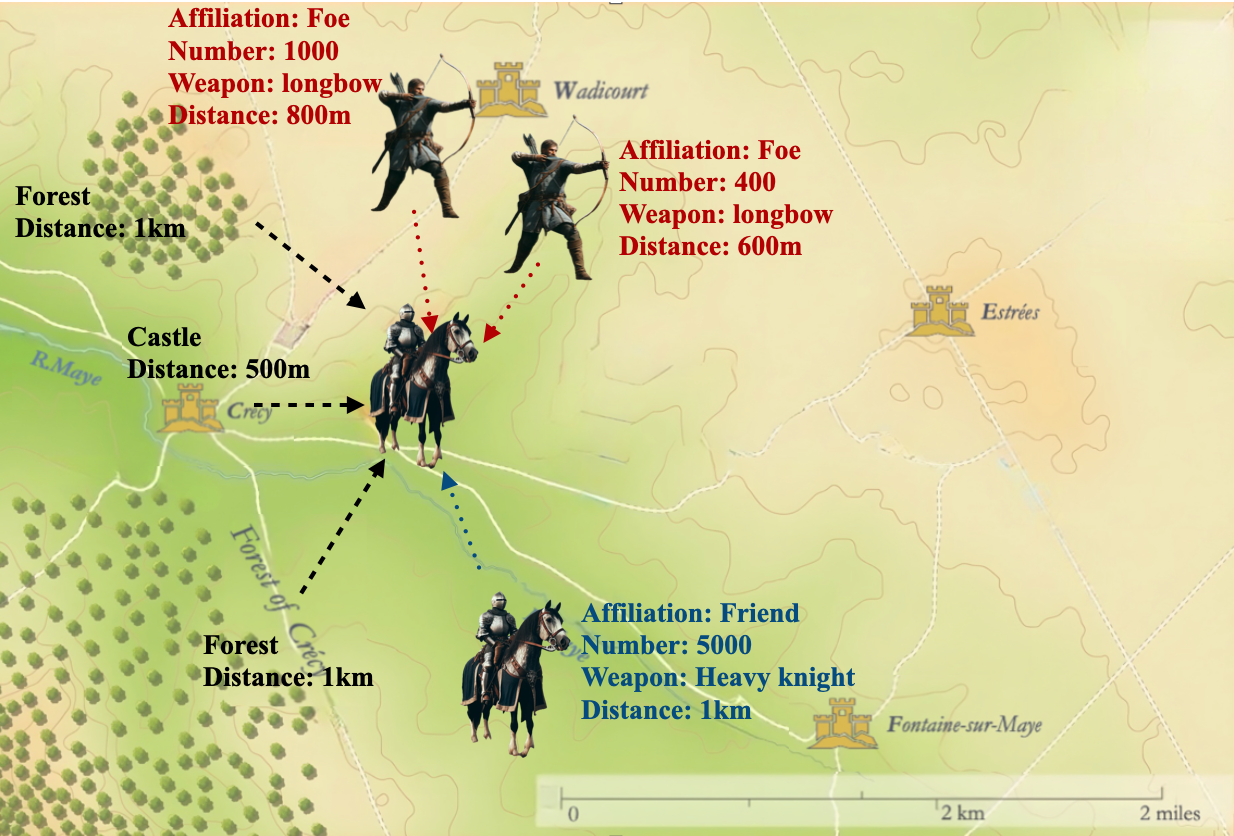}
    \caption{Observation based on Map Information.}
    \label{fig:observation}
\end{figure}

In line with the behavior of all agents, the initial step prior to planning or executing any actions involves observing the overall environment information. To achieve this, we adopt a multi-modal approach that amalgamates both textual and visual representations of the environment. A visual map is presented to the agents, depicting their starting locations and specific places of interest, such as villages. In addition, the agents are provided with corresponding textual descriptions of the coordinates in their prompts. Subsequently, each agent proceeds to collect data in two primary aspects:
\begin{enumerate}
    \item Geographic landscapes: Geographic information offers a macroscopic view, providing agents with critical insights for strategic planning and navigation.
    \item Nearby agents: Agents information including affiliation (friend or foe), current actions, precise positioning (distance and bearing), and detailed agent profiles including the remaining number of soldiers and their composition.
\end{enumerate}

Together, these observation streams form a comprehensive situational awareness framework, crucial for the agents' adaptive responses and operational effectiveness in diverse settings. An example is presented in Figure \ref{fig:observation}.

To simulate the agents' limited field of view, each agent's prompt only includes the coordinates of agents and places within a 10k-meter range. All agents then calculate the relative distance of other agents and places within their 10k-meter sight range using both textual and visual information. It is important to note that if obstacles like forests or villages come into an agent's sight, the agent will not be able to see through these obstacles. By combining textual and visual descriptions, our multi-modal approach enables agents to interact effectively with their surroundings while maintaining a realistic representation of their limited field of view. 

If sub-agents collect information and communicate with the parent agent, the parent agent's knowledge can extend beyond its immediate sight range. This information exchange allows the parent agent to make more informed decisions based on the broader context provided by its sub-agents, enabling more effective coordination and strategic planning. In such cases, the parent agent's awareness of the battlefield is not strictly limited to its own 1,000-meter sight range but can be expanded through the information gathered by its sub-agents. This extended knowledge can include details about enemy positions, terrain features, or other relevant factors that might influence the parent agent's decision-making process. As a result, the dynamic multi-agent system can better adapt to the complex and evolving battlefield environment by leveraging the combined knowledge of its constituent agents.

\subsubsection{Action Planning}

At each discrete time point, an agent has the ability to choose from a multitude of potential actions. In this part, we will outline four common actions that agents typically engage in: location movement, dynamic agent structure, interaction with the landscape, and interaction with other agents. These actions encompass a range of strategic and tactical considerations that agents must take into account when making decisions in the context of the battlefield.

\paragraph{Location Movement} In the context of location movement, an agent possesses the capability to traverse to a different location for strategic purposes. This may involve moving closer to enemy agents to initiate an attack, or distancing itself from potential threats. In terms of the mechanics of location movement, the agent will generate the coordinates of its intended final destination, which it aims to reach within the subsequent 15-minute timeframe.

The following represents an illustrative output of a location movement action for an agent belonging to England with the ID \textsc{ARMY-a606969b}. In this scenario, the agent moves from its current location, represented by the coordinates [95, -55], to a new location with coordinates [100, -50]. During the course of this movement, the agent sustains losses of 30 soldiers. 
\begin{lstlisting}[language=HTML, caption=An example output of location movement action]
{
identity: England,
id: ARMY-a606969b,
action_description: Reposition Forces Reposition to [100, -50] to utilize ForestF for cover and longbow deployment, supporting nearby friendly units and executing flanking maneuvers,
current_location: [95, -55],
target_location: [100, -50],
remaining_number: 170,
original_number: 200,
lost_number: 30
}
\end{lstlisting}

\paragraph{Dynamic Agent Structure}
The battlefield environment is highly dynamic and fluid, with a multitude of situations arising unpredictably. To address this complexity, we propose a dynamic agent structure \cite{liu2023dynamic, han2024llm} that enables agents to adapt their organizational configurations according to the current situation. Our proposed dynamic agent structure supports several adaptive mechanisms, as shown in Figure \ref{fig:agent_structure}:

\begin{figure}[!ht]
    \centering
    \includegraphics[scale=0.32]{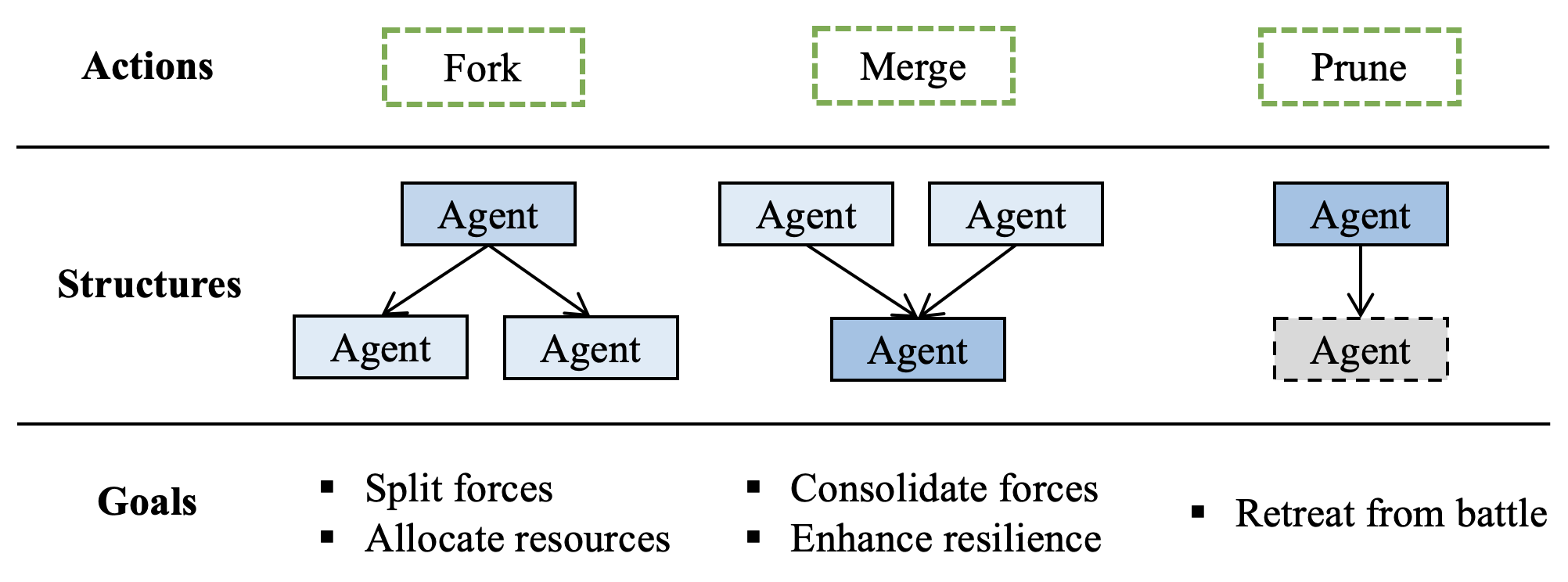}
    \caption{Dynamic agent structure.}
    \label{fig:agent_structure}
\end{figure}

\begin{enumerate}
    \item Fork: An agent may decide to fork another autonomous agent for a specific task, effectively splitting its forces and allocating resources to address multiple objectives simultaneously.
    \item Merge: In scenarios where an agent is under significant pressure but chooses to continue fighting, it may merge with the closest allied agent to consolidate forces and enhance its resilience.
    \item Prune: In cases where an agent is overwhelmed or retreats from the battlefield, the dynamic agent structure accommodates this change by pruning the agent from the active force.
\end{enumerate}

This dynamic agent structure allows for flexible and adaptive decision-making in response to the ever-changing conditions of the battlefield. For instance, an agent may fork a sub-agent by allocating a portion of its soldiers to create an autonomous agent responsible for collecting information about the surroundings and communicating it back. Additionally, the agent may dispatch autonomous sub-agents to focus on handling specific segments of the incoming enemy forces based on the situation's requirements. In cases where agents lack the necessary strength to address the current situation, they may choose to retreat from the map or merge with other agents to bolster their combined power. Therefore, our project introduces a dynamic multi-agent system in which the structure and number of agents adapt according to the needs of the given situation. This adaptive approach allows the system to effectively respond to the diverse and unpredictable challenges that arise in the battlefield environment.

The following represents an example of a forking action, where the current agent with the ID \textsc{ARMY-4c9b5218} decides to split its army into two distinct agents. The first new agent consists of 500 soldiers, all of whom are light cavalry, and is assigned the mission of initiating a skirmish at the location of [25, -75] with an enemy agent with the ID \textsc{ARMY-ac57a3f0}. This new agent is assigned the ID \textsc{ARMY-ac57a3f0}. The second new agent consists of 4,000 soldiers, all of whom are heavy cavalry, and is assigned the mission of rallying troops at the location of [25, -75]. This new agent is assigned the ID \textsc{ARMY-70b7fa20}. This example demonstrates the dynamic nature of agent structure within the emulation sandbox, where agents can split and form new agents with distinct properties and missions.
\begin{lstlisting}[language=HTML, caption=An example of forking of agents]
Current agent: 
{
identity: France,
agent_id: ARMY-4c9b5218,  
troop_number: 6000, 
current_location: [20, -70]
}

Given the current battlefield situation, where the enemy is deploying longbows and fortifying positions to counter our cavalry charge, we need to adapt our strategy. Our initial mission was to charge cavalry, but the enemy's longbows pose a significant threat to our heavily armored knights, especially if we charge without dealing with them first.

## Thought for Action:
1. We need to disrupt the enemy's longbowmen, who are currently thebiggest threat to our cavalry.
2. We should avoid charging directly into fortified positions or where longbowmen are deployed.
3. Coordinating with friendly forces is crucial to create a unified front and increase the effectiveness of our actions.
4. Considering the terrain and the positioning of our forces, we should aim to flank the enemy where possible.
5. We must maintain high morale among our troops to ensure the effectiveness of our charge.

## The Actions to Perform:
- **Reposition Forces**: To avoid a head-on charge into the enemy's longbows, we will reposition our forces to prepare for a flanking maneuver.

fork agents: 
[
{
subAgent_NextActionType: Initiate Skirmish,
troopType: Light Cavalry,
deployedNum: 500,
target_position: [25, -75],
target_agent_id: ARMY-75b59d12,
agent_id: ARMY-ac57a3f0,
remarks: Engaging enemy longbowmen to disrupt their formation and 
reduce their effectiveness.
},

{
subAgent_NextActionType: Rally Troops,
troopType: Heavy Cavalry,
deployedNum: 4000,
target_position: [20, -70],
target_agent_id: None,
agent_id: ARMY-70b7fa20,
remarks: Rallying the main force to maintain high morale and prepare for the charge.
}
]
\end{lstlisting}

In Section 3.2, the general profile of commanding agents has been delineated. However, we can see from the dynamic agent structure here that agents are dynamic entities in our sandbox, and within a single country's army, there may be numerous distinct agents, each engaged in different tasks. Therefore, in addition to the general information inherited from the overall commanding agent profile as defined in Section 3.2, each agent possesses more granular and unique information, as defined by the following dynamic properties:

\begin{enumerate}
\item Initial mission assigned when being created
\item Current location represented by coordinates
\item The number of soldiers at its disposal
\item The type of soldiers under its command
\end{enumerate}

These properties are subject to evolution over time. For instance, the number of soldiers associated with an agent may fluctuate as a result of soldiers joining the agent, thereby increasing its forces, or from soldiers being killed or wounded in battle, leading to a decrease in its forces. The current location of the agent may also change as it navigates the battlefield, and its initial mission may adapt in response to shifting circumstances and strategic considerations.

\begin{figure}[!ht]
    \centering
    \includegraphics[width=\textwidth]{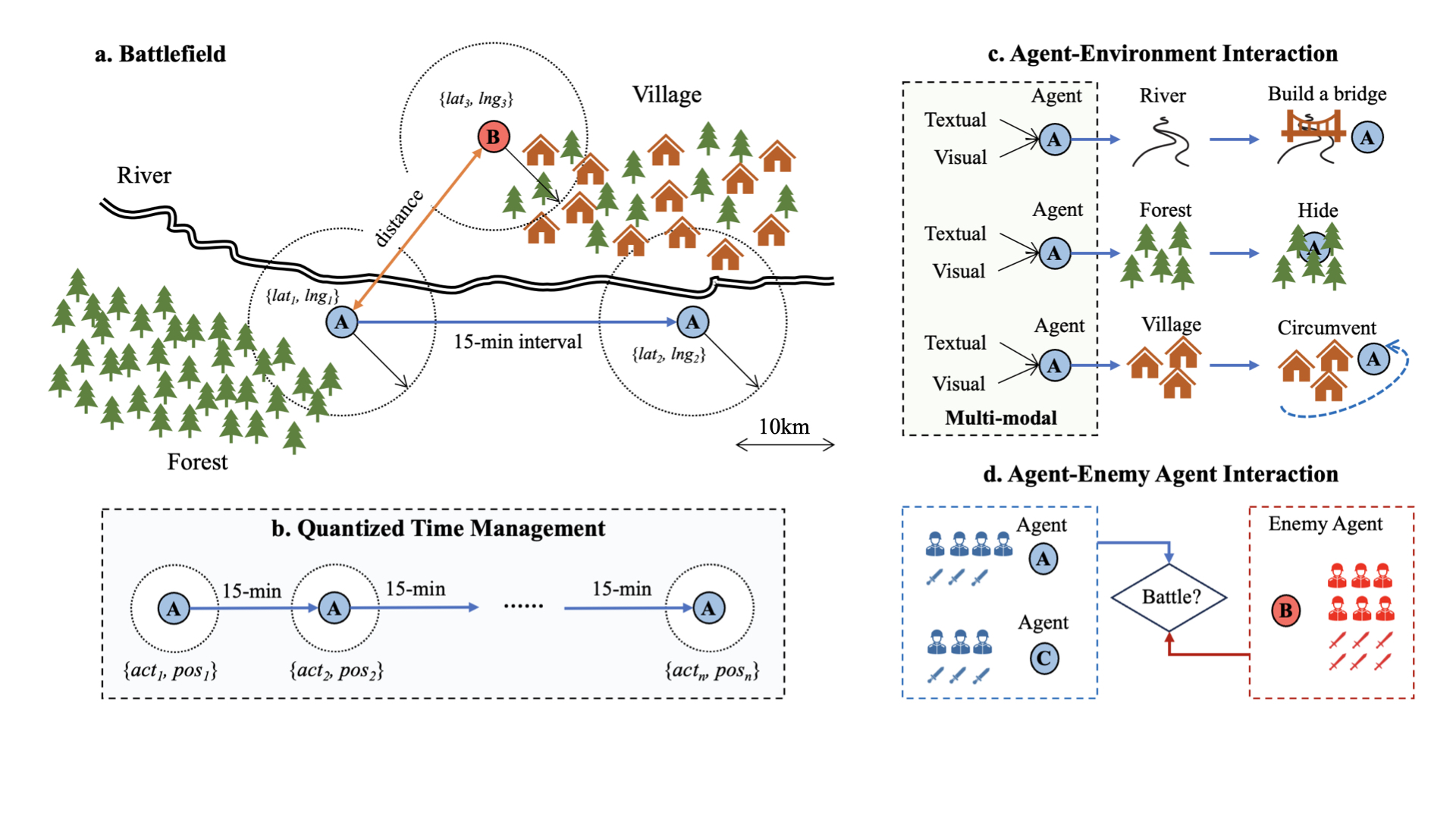}
    \caption{Battlefield interaction (a) Battlefield environment, (2) Quantized time management, (c) Agent-environment interaction, and (d) Agent and enemy agent interaction.}
    \label{fig:battle_field}
\end{figure}

\paragraph{Interaction with Landscape Environment}
To accurately emulate battle dynamics, it is crucial for agents to be able to interact with the physical surroundings as shown in Figure \ref{fig:battle_field} (c), such as rivers, forests, villages, and other features. For example, when encountering a river, agents may build a bridge to cross it; when encountering a forest, agents might choose to hide within it to ambush enemies; and when encountering a village, agents could decide to circumvent it. To facilitate these interactions, it is essential to maintain a relative distance between agents and specific locations on the map, as well as between agents themselves. 

The following represents an example of an agent utilizing the natural cover provided by a forest. In this scenario, the agent strategically positions itself within the forest to gain a tactical advantage, such as concealment from enemy agents or protection from ranged attacks.
\begin{lstlisting}[language=HTML, caption=An example of an agent interacting with the landscape of forest]
{
identity: England,
id: ARMY-53c7a137,
action_description: Fortify Position Constructing defenses in ForestF to utilize the natural cover against cavalry charges and to bolster the position of our longbowmen,
location: [100, -50],
remaining_number: 300,
original_number: 300,
lost_number: 0
}
\end{lstlisting}

\paragraph{Interaction with Other Agents}
Given the observation agents make about their surrounding situations, agents will make decisions regarding whether and when to engage in interactions with other agents, particularly those identified as enemies, as depicted in Figure \ref{fig:battle_field} (d). The specific nature and timing of these interactions are not predetermined; rather, they are initiated by the agents themselves. For instance, when an enemy agent is within close proximity, an agent may opt to engage in combat or launch an attack. The outcome of these interactions between agents is contingent upon various factors, such as the number of soldiers at their disposal and the types of weapons they possess.

The following represents two examples of offensive interactions between two enemy agents. In the first example, an agent executes a flanking maneuver against another agent, as shown in the code snippet below:
\begin{lstlisting}[language=HTML, caption=An example of an agent interacting with a target agent with attacks]
{
identity: France,
id: ARMY-d2ff280c,
action_description: Execute Flanking Maneuvers Flanking enemy unit 'ARMY- 9418a275' in the midst of a Tactical Retreat to destabilize their fortification efforts,
location: [10, -100],
remaining_number: 8200,
original_number: 12000,
lost_number: 3800
}
\end{lstlisting}

In this scenario, the agent with the ID \textsc{ARMY-d2ff280c}, representing France, executes a flanking maneuver against the enemy unit with the ID \textsc{ARMY-9418a275}. The maneuver is intended to destabilize the enemy's fortification efforts while they are in the midst of a tactical retreat. The agent's current location is represented by the coordinates [10, -100], and it has a remaining force of 8,200 soldiers out of an original force of 12,000, having lost 3,800 soldiers.

The second example involves an agent ambushing an enemy cavalry unit, as shown in the code snippet below:

\begin{lstlisting}[language=HTML, caption=Another example of an agent interacting with a target agent with attacks]
{
identity: England,
id: ARMY-2508af97,
action_description: Ambush Enemy Proactively engaging enemy cavalry at [83.8, -17.4] to disrupt their maneuvers and prevent them from supporting their troops,
location: [20, -25],
remaining_number: 600,
original_number: 600,
lost_number: 0
}
\end{lstlisting}

Notice that there are numerous actions that agents can undertake even when they are not directly interacting with each other. These actions may include fortifying their position, rallying troops, and other similar activities that contribute to their overall strategic advantage on the battlefield.

\subsubsection{Casualty Evaluation by Observer}
In the event that one agent initiates an aggressive action towards another, hereafter referred to as the target agent, both parties may sustain corresponding casualty losses. The loss is evaluated by an objective evaluator supported by GPT-4, which can be seen as an observer. The observer determines the casualties based on several factors:

\begin{enumerate}
\item Current profile information from the agents, including their force size, force composition, command architecture, and location.
\item The actions undertaken by the agents, including the action name and a more detailed description of the action generated alongside the action name by the agent. For example, ``Deploy Longbows: Deploying longbows in coordination with nearby friendly forces to initiate a skirmish against the nearest enemy cavalry unit and disrupt their advance.''
\item The location and relative distance between the agents, as well as relevant landscape information surrounding them. This information is used to assess the tactical advantages or disadvantages of the agents' positions.
\item Objective information about the specific weapon utilized, including performance metrics such as the range and accuracy of the weapon. This information is obtained from reputable sources such as Wikipedia.
\end{enumerate}

In order to improve the accuracy and reliability of casualty assessments, it is recommended that future iterations of the emulation sandbox incorporate an \textit{expert system} with a more comprehensive understanding of the weapons involved. Such a system would be able to provide more nuanced and accurate evaluations of casualties based on a deeper understanding of the capabilities and limitations of different weapons, as well as the tactics and strategies employed by the agents.

\subsubsection{Agent Profile Update}
The profile information of each agent, encompassing force size, force composition, command architecture, and location, is dynamically updated based on the actions undertaken by the agent at the current time. The profile information is updated for every quantized time period.

The factors taken into account to update the agent profile include the actions undertaken by the agent, their interactions with other agents, and any movement that occurs. Specifically, the force size of an agent is determined by three key factors: the change in casualty numbers, the forking of the agent, or the merging of other agents. An agent can decide whether to fork more agents or to merge with other agents, which will result in an increase or decrease in force size, respectively. The location of an agent is contingent upon their movement. If a movement action is executed, the agent's location is updated accordingly. The updates to the agent profile ensure that the emulation sandbox accurately reflects the current state of the battlefield and the evolving dynamics of the conflict.

\subsubsection{Historical Action Trajectory}
For all agents, once decisions regarding actions have been made at each quantized time period, these actions are subsequently recorded into a historical action trajectory, which is then incorporated into the general prompt. As a result, all future decision-making processes will be informed not only by the current environmental information but also by the historical trajectory of actions that have been previously undertaken. This approach enables agents to make informed decisions that are grounded in both the current context and the historical record of actions on the battlefield.

\subsection{Emulation of Single Soldier}
Our sandbox also simulates the experiences of individual soldiers and reflects their personal perspectives. Throughout the emulation, these soldier agents document their experiences and emotions based on their current actions, previous action trajectories, and any wounds they may have sustained from enemy attacks on the agent to which they belong. This process provides valuable insights into the personal aspects of warfare that complement the higher-level strategic decision-making processes carried out by the commanding agents.

By adopting a multi-layered approach, our emulation is able to capture both the macro-level strategic dynamics and the micro-level personal experiences of the battlefield. This results in a more comprehensive representation of the complexities of war, encompassing both the broader strategic considerations and the individual experiences of soldiers on the ground.

\section{Experiment}
The primary objective of these experiments is to investigate the extent to which agents based on LLMs and VLMs can reasonably simulate historical battles, which are characterized by a high degree of complexity and dynamism. Specifically, we aim to evaluate the ability of agents to effectively navigate and adapt to the rapidly evolving and unpredictable situations that typically arise during battles. By doing so, we hope to gain insights into the potential of LLMs and VLMs as tools for simulating and analyzing historical conflicts.

We conduct experiments on 4 distinct historical scenarios, namely the Battle of Crécy, the Battle of Agincourt, the Battle of Falkirk, and the Battle of Poitiers. The experiments are performed using 3 strong language models and vision-language models: Claude-3-opus \cite{claude3}, GPT-4-1106-preview \cite{achiam2023gpt}, and GPT-4-vision \cite{gpt4v}. For each scenario and each language model, we execute the emulation 5 times using the same setting within a sandbox environment to account to randomness, continuing until the casualty figures for both armies converge, or in other words, reach a state of stability.

\subsection{Evaluation Metrics}

As historical battles often lack comprehensive records and documentation, and are typically characterized by unpredictable events that are challenging to replicate in an emulation, our evaluation methodology is divided into three distinct components.

\begin{table}[!ht]
\centering
\begin{tabular}{p{4.5cm}|p{9cm}}
\toprule
\textbf{Evaluation aspect} & \textbf{Description} \\
\hline
Final battle casualty & Comparison with historical data, focusing on the final casualty figures for both armies \\
\hline
Human analysis on location movement & Assessment of the dynamic structure of agents and their movement on the battlefield as a whole \\
\hline
Human analysis of agent action & Evaluation of the reasonableness of the actions conducted by the agents. \\
\bottomrule
\end{tabular}
\caption{Three aspects of evaluation and demonstration.}
\end{table}

\paragraph{Battle Final Casualty} The first aspect focuses on the final battle casualty, where we compare the simulated casualty figures with historical data. Specifically, we evaluate the casualty of each army for every quantized time period, recording the casualty for each time for both armies. We then compute the mean and variance at each time and compare the final result with historical data to determine the degree of alignment between the emulation results and the historical record. This approach allows us to assess the accuracy and reliability of the emulation sandbox in replicating historical battles.

\paragraph{Human Analysis on Location Movement} The second part of the evaluation involves a human analysis of the agents' behavior on the battlefield, specifically their location movements and dynamic structure. To facilitate an intuitive assessment of whether the agents can rationally interact with the environment and make decisions about where to go, we present images of the action dynamics and the positions of the agents across time. This allows us to evaluate the agents' ability to adapt to changing circumstances and execute historically plausible strategies, providing insights into the effectiveness of the emulation sandbox in simulating realistic battle scenarios.

\paragraph{Human Analysis on Agent Action} The third aspect of the evaluation involves checking the actions taken by the agents. To provide a clear and understandable representation, we present a sequence of diagrams that track the action trajectories taken by specifically two agents. 

Although the latter two aspects are challenging to quantitatively evaluate, hopefully by presenting the agents' behavior and actions in a visual format, we can offer insights into the agents' decision-making processes and their ability to execute historically plausible strategies.

\subsection{First Aspect: Battle Result Performance}

\begin{figure}[!ht]
    \includegraphics[scale=0.185]{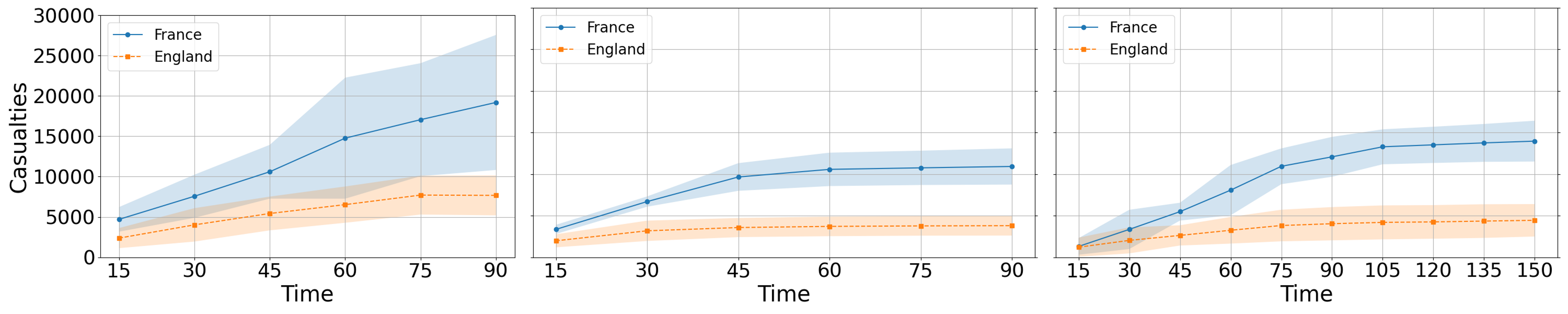}
    \caption{Emulated casualty results for Battle of Crécy.}
    \label{fig:Crecy}
\end{figure}

\begin{figure}[!ht]
    \centering
    \includegraphics[scale=0.22]{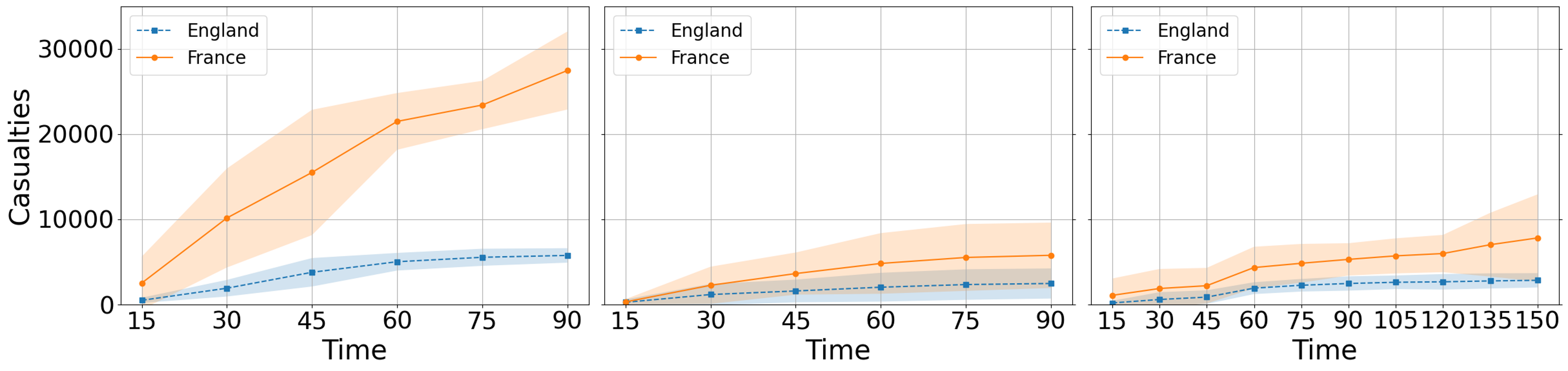}
    \caption{Emulated casualty results for Battle of Agincourt.}
    \label{fig:Agincourt}
\end{figure}

\begin{figure}[!ht]
    \centering
    \includegraphics[scale=0.22]{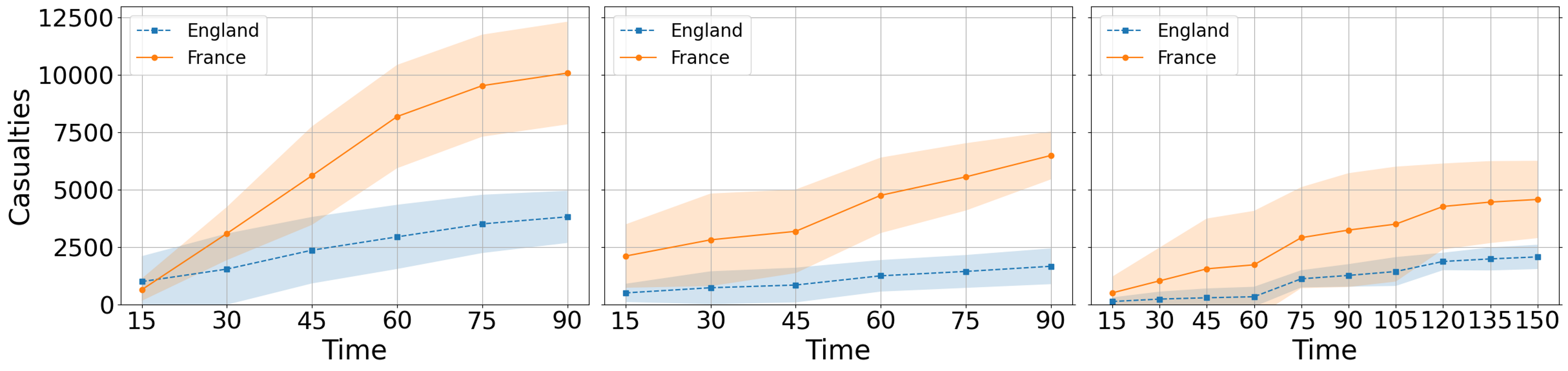}
    \caption{Emulated casualty results for Battle of Poitiers.}
    \label{fig:Poitiers}
\end{figure}

\begin{figure}[!ht]
    \centering
    \includegraphics[scale=0.22]{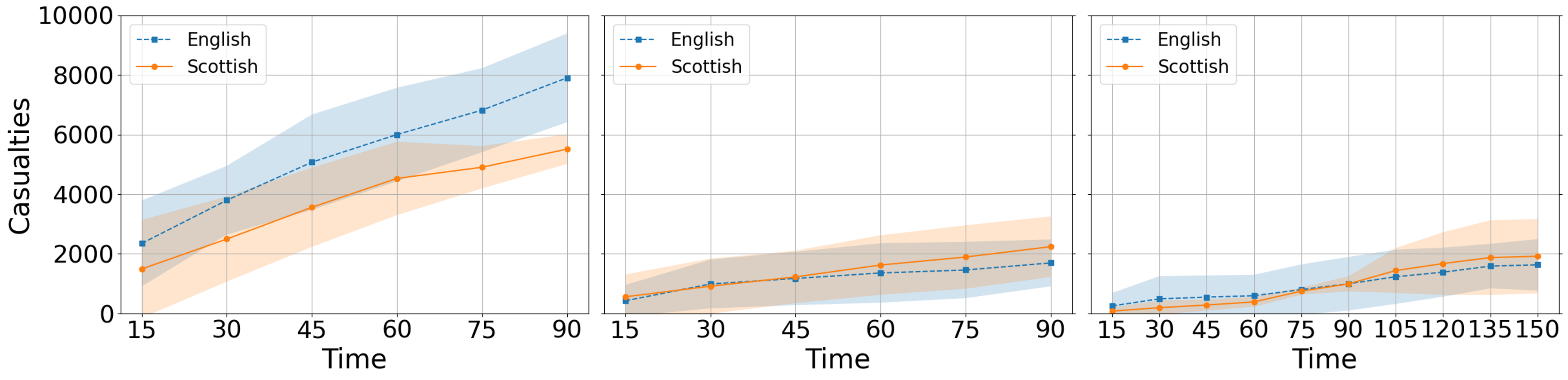}
    \caption{Emulated casualty results for Battle of Falkirk.}
    \label{fig:Falkirk}
\end{figure}

In this section, we present the results of the first aspect of the evaluation, which focuses on comparing the simulated casualty figures with historical data. Figures \ref{fig:Crecy}, Figures \ref{fig:Agincourt}, \ref{fig:Poitiers}, and \ref{fig:Falkirk} illustrate the results of simulating the four battles using the three LLM-based and VLM-based agents. 

For each battle emulation, we utilize 3 different backbone models to generate results. The results are presented in 3 separate images, with the leftmost figure showing the results from the Claude-3 simulations, the middle figure showing the results from the GPT-4 simulations, and the rightmost figure showing the results from the GPT-4-vision simulations. This allows us to compare the performance of the different language models in replicating historical battles and to evaluate the impact of the choice of language model on the accuracy and reliability of the simulation results. For each image, the x-axis represents the process of time in emulation, recorded in minutes, while the y-axis represents the casualty figures for armies belonging to the two countries. The casualty of each army is represented by a mean line and a variance band, where the mean and variance are computed based on the 5 simulations run in the sandbox in the current setting. This allows us to assess the accuracy and reliability of the emulation sandbox in replicating historical battles and to evaluate the impact of the choice of language model on the simulation results.

Overall, the results indicate that Claude-3 predicts a significantly higher casualty rate compared to GPT-4 and GPT-4-vision. This discrepancy suggests that the choice of backbone language model can have a significant impact on the accuracy and reliability of the emulation sandbox in replicating historical battles.

\paragraph{Battle of Crécy}
In the emulated Battle of Crécy, the results indicate that all three models predict a higher casualty rate for the French soldiers compared to the English soldiers, which aligns with historical records.  Specifically, Claude-3's prediction suggests that the final casualty figure for the French army is approximately 2.4 times that of the English army. However, the prediction is characterized by high variance, which increases as the emulation progresses, suggesting a high degree of randomness in the emulation process. In contrast, GPT-4's prediction suggests that the final casualty figure for the French army is approximately 2.75 times that of the English army, with a relatively low variance, indicating a stable emulation process. Similarly, GPT-4-vision's prediction suggests that the final casualty figure for the French army is approximately 3.5 times that of the English army, with an acceptable level of variance.

Overall, the results suggest that all three agents are able to simulate the casualty figures for the Battle of Crécy with varying degrees of accuracy and stability. While Claude-3's prediction is characterized by high variance, GPT-4 and GPT-4-vision's predictions are more stable and align more closely with historical records.

\paragraph{Battle of Agincourt}
In the emulated Battle of Agincourt, the results indicate that all three models predict a higher casualty rate for the French soldiers compared to the English soldiers, which aligns with historical records. Based on historical documentation, only a few hundred died in the English army and about 4,000 to 10,000 died in the French army. Among the three models, GPT-4 and GPT-4-vision predict a closer casualty result, though the casualty of the English is still much higher than historical fact, but the casualty in French are relatively similar. The variance in the predictions is within an acceptable range. However, Claude-3 predicts a much higher casualty for both English and French armies, with the French army almost dying out, which is not very reasonable.

Overall, the results suggest that GPT-4 and GPT-4-vision can provide a relatively reasonable result for the Battle of Agincourt, while Claude-3's predictions are less accurate. This highlights the importance of selecting an appropriate language model for the emulation sandbox to achieve historically plausible results.

\paragraph{Battle of Poitiers}
In the emulated Battle of Poitiers, the results show a much higher casualty for the French than the English army, conforming to historical documentation: only around 40 soldiers died in this battle in the English army, but more than 4,500 men-at-arms were killed or captured. GPT-4 and GPT-4-vision predict an acceptable result for the French army, which is around 4,000 to 6,000, but too high for the English army, which is close to 2,000 for both models. Claude-3 again predicts too high casualty for both English and French armies.

Overall, the results suggest that GPT-4 and GPT-4-vision can provide a relatively reasonable result for the Battle of Poitiers, while Claude-3's predictions are less accurate.

\paragraph{Battle of Falkirk}
In the emulated Battle of Falkirk, the results show a similar casualty for the English army and the Scottish army, conforming to historical documentation: around 2,000 soldiers died in this battle in both the English and Scottish armies. GPT-4 and GPT-4-vision predict a very close result, with the casualty rate for both English and Scottish armies around 2,000. Again, Claude-3 predicts too high casualty for both armies, with English around 8,000 and Scottish around 6,000.

Overall, the results suggest that GPT-4 and GPT-4-vision can provide a relatively reasonable result for the Battle of Falkirk, while Claude-3's predictions are less accurate.

\subsection{Spatial Movement Result Performance}
\begin{figure}[!ht]
    \centering
    \includegraphics[scale=0.39]{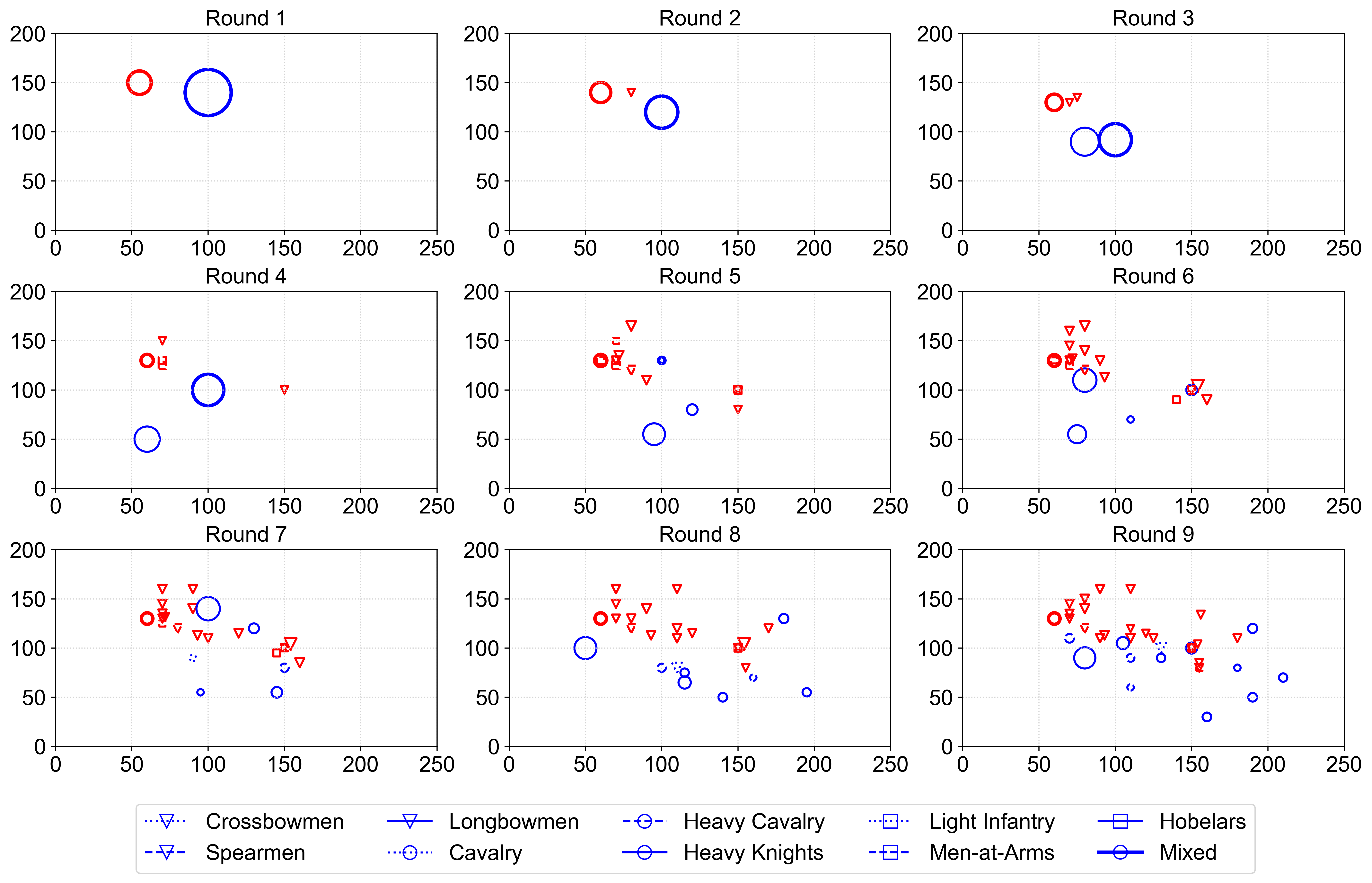}
    \caption{All agent movement and dynamic agent structure on battlefield.}
    \label{fig:movement_example}
\end{figure}

Figure. \ref{fig:movement_example} illustrates the general agent location dynamics of a single emulation of the Battle of Crécy using GPT-4. The English army is represented by red symbols, while the French army is represented by blue symbols. The sizes of the symbols are normalized to correspond to the number of soldiers contained in each agent. Different line type

At a glance, we can observe that as the emulation progresses, both armies are gradually split into smaller teams, especially the English army. Notably, some of the longbowmen tend to maintain a safe distance from the enemy for extended periods, using their longbows to inflict casualties from afar. As time progresses, the advantage of the French army's larger number of soldiers is diminishing over time, particularly in the case of the heavy cavalry and heavy knights. This is likely due to the effectiveness of the English longbowmen in inflicting casualties from a safe distance, as well as the challenging terrain of the battlefield, which made it difficult for the heavily armored French knights to maneuver effectively.

To further evaluate the performance of the LLMs and VLMs in simulating historical battles, we can examine the paths taken by individual agents over time. This can provide insights into whether these models have a good sense of distance and can make reasonable decisions based on the overall environment.

\subsection{Agent Action throughout Emulation}

\begin{figure}[!ht]
    \centering
    \includegraphics[scale=0.63]{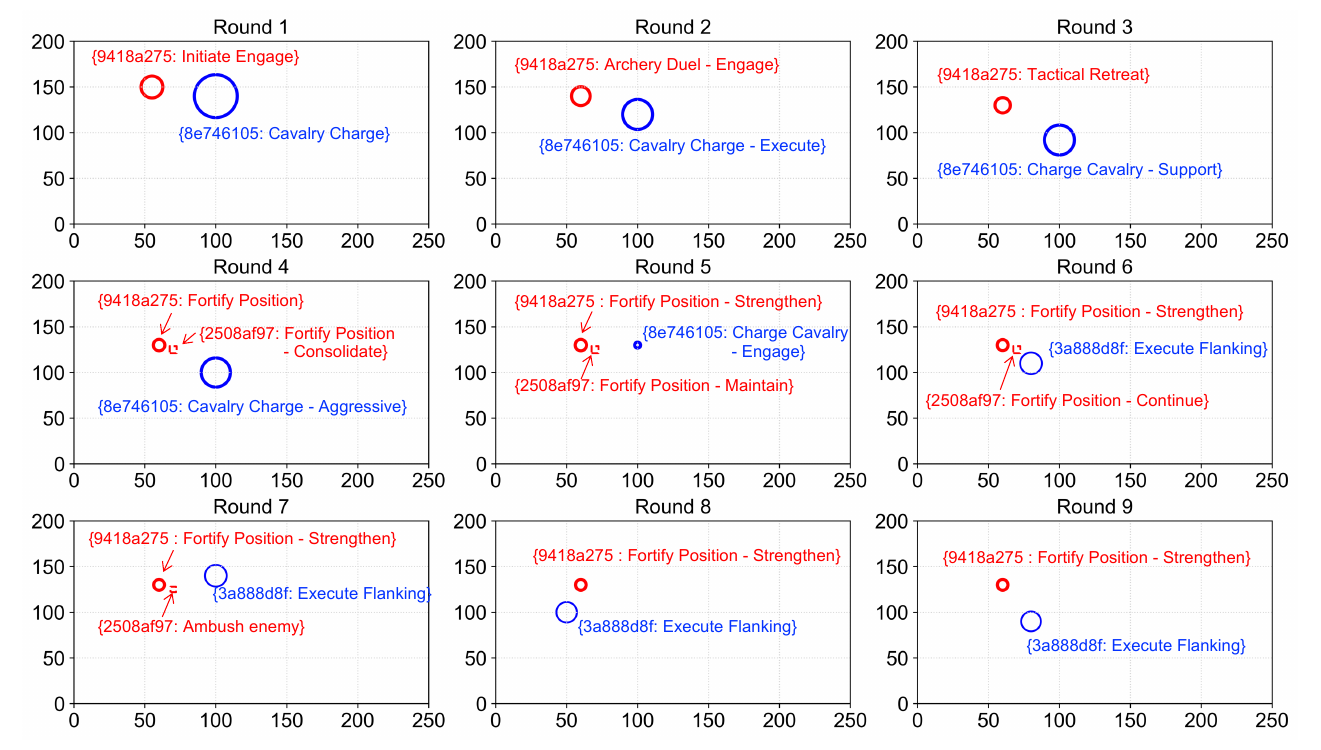}
    \caption{Agent action tracker over time.}
    \label{fig:agent_track}
\end{figure}

Figure \ref{fig:agent_track} provides an illustrative example of the actions undertaken by two agents, one representing a part of the army belonging to England and the other representing a part of the army belonging to France, throughout the entire emulation time. The English agent's cautious approach is reflected in its movements and actions, while the French agent's aggressive strategy is evident in its frequent attacks and resulting losses. This example provides a reasonable representation of how historical battles may have unfolded.

\subsection{Experiment Summary}
The evaluation results of the emulation sandbox for historical battles indicate that the choice of language model can have a significant impact on the accuracy and reliability of the simulation. Specifically, GPT-4 and GPT-4-vision have shown to provide relatively reasonable results in terms of casualty figures for both armies in the emulated battles of Agincourt, Poitiers, and Falkirk, while Claude-3 has consistently predicted much higher casualty rates than historical records.

In terms of the strategies and tactical maneuvers employed by the agents, the results suggest that the agents can adapt to changing circumstances and execute historically plausible strategies. However, further analysis is needed to assess the agents' ability to rationally interact with the environment and make decisions about where to go.

Overall, the evaluation methodology employed in this study, which combines a comparison with historical data and a human analysis of the agents' behavior, has proven to be effective in assessing the accuracy and reliability of the emulation sandbox. However, further refinement is needed to improve the accuracy of the predictions and the agents' decision-making capabilities.

\section{Soldier Agent Experience}
The experiences of individual soldiers are recorded for each quantized time interval during the simulation. Each soldier agent generates a document based on their unique background and experiences. Therefore, for a simulation involving 30 soldier agents, a total of 30 documents are produced, each containing multiple episodes.

To provide an overview of the content of these documents, we utilize word frequency analysis and present the results using a word cloud. This involves merging all documents generated by soldier agents from the same country, regardless of the quantized time interval. The resulting word cloud offers a visual representation of the most frequently occurring words and phrases in the documents, providing insights into the experiences and perspectives of the soldier agents.

To generate the word cloud, we first preprocess the text data by removing words with specific part-of-speech tags, such as those belonging to the tag group [RB, MD, IN, CD, DT, NNS, PRP, NNS, FW]. We also remove functional words that do not convey meaningful information. Additionally, we filter out high-frequency words that do not provide significant insights, such as ``think'', ``feel'', ``battle'', and ``war''. This preprocessing step helps to ensure that the resulting word cloud accurately reflects the most salient themes and concepts present in the soldier agents' documents.

\begin{minipage}[t]{0.5\textwidth}
\includegraphics[width=\textwidth]{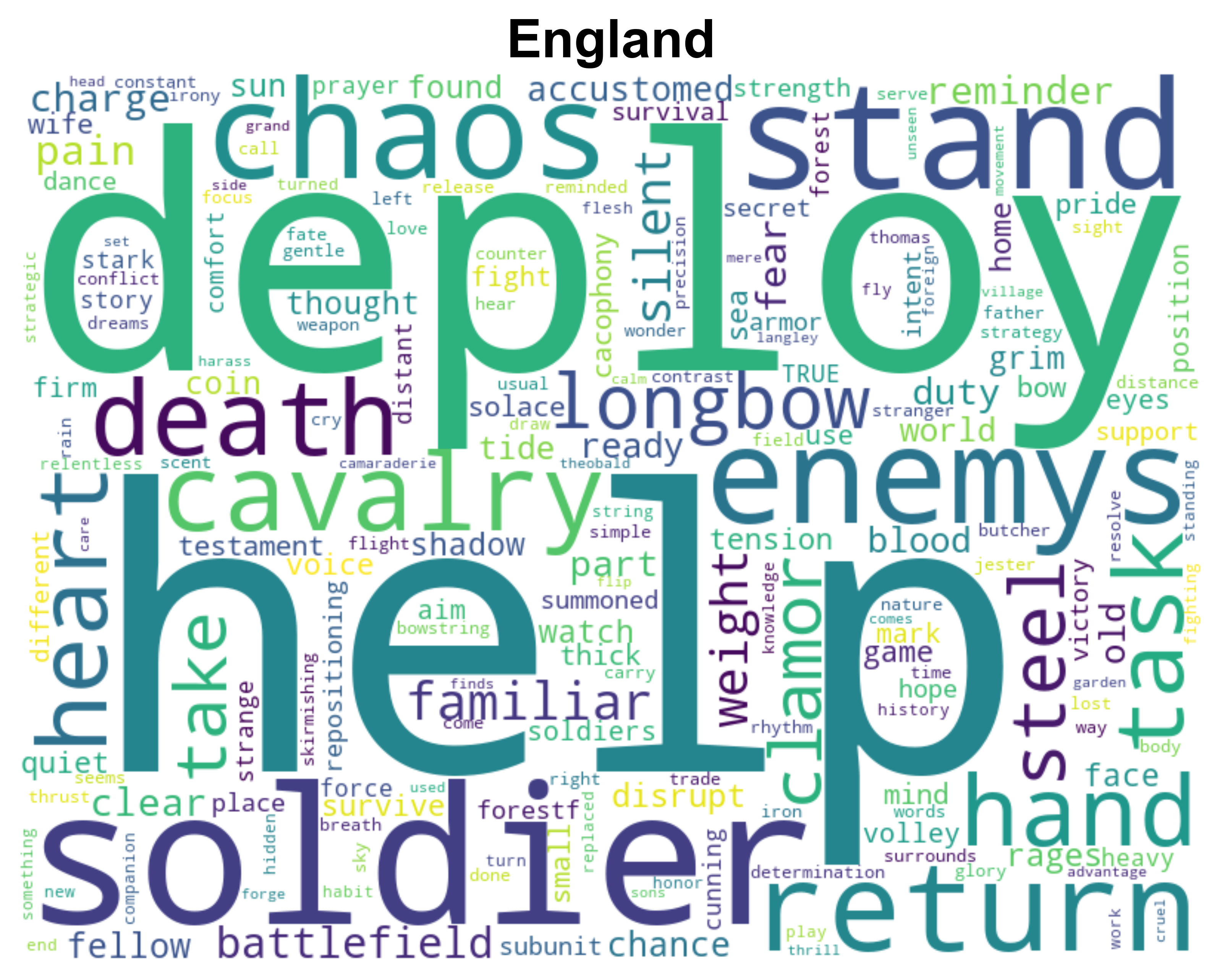}
\end{minipage}
\begin{minipage}[t]{0.5\textwidth}
\includegraphics[width=\textwidth]{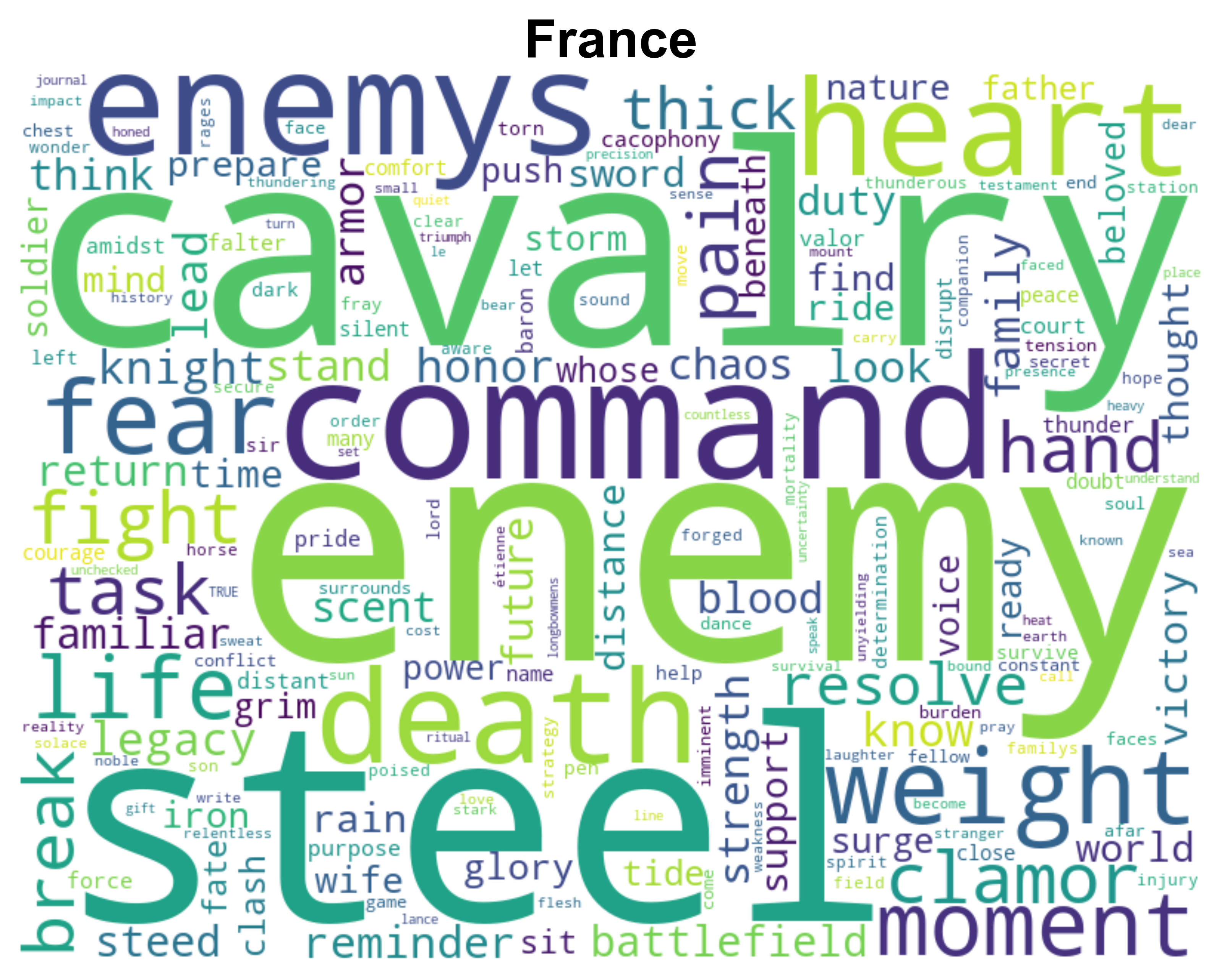}
\end{minipage}

The above two figures present word clouds generated from the documents written by soldier agents in the English and French armies during the Battle of Crécy. Despite being the winning army, the English army's word cloud reveals a high frequency of words such as ``death'', ``chaos'', and ``return'', indicating the intense and chaotic nature of the battle. On the other hand, the French army's word cloud shows a high frequency of words such as ``death'' and ``fear'', reflecting the fear and uncertainty experienced by the soldiers. These word clouds provide a visual representation of the experiences and emotions of the soldier agents in both armies during the battle.

\section{Conclusion and Future Work}
\paragraph{Conclusion}
In this study, we have demonstrated the potential of LLM and VLM to support highly complex and dynamic simulations of historical battles. Our emulation sandbox provides a comprehensive evaluation of the simulated battles, including a comparison of casualty figures with historical data and a human analysis of the strategies and tactical maneuvers employed by both armies. Our approach also presents the individual experiences of soldiers on the battlefield using soldier agents, providing valuable insights into the personal aspects of warfare that complement the higher-level strategic decision-making processes carried out by the commanding agents.

We believe that our work can also provide new pedagogical methods for students and researchers interested in historical analysis. By simulating historical battles and presenting the results in an interactive and intuitive way, students can gain a deeper understanding of the complexities and dynamics of warfare. Moreover, our approach can be used to support research in various fields, such as military history, AI, and game theory.

\paragraph{Future Work}
Our study has demonstrated the potential of LLM and VLM to support complex and dynamic simulations of historical battles. However, there is still room for improvement and expansion of our approach.

Firstly, we aim to develop \textbf{additional evaluation metrics} for such dynamic simulations to establish their effectiveness more comprehensively. This will enable us to better assess the accuracy and reliability of the simulation results and identify areas for improvement.

Secondly, we plan to extend our approach to \textbf{simulate different types of battles beyond barpitite medieval battles}. This will allow us to evaluate the versatility of our approach and its applicability to a wider range of historical battles.

Thirdly, we aim to \textbf{incorporate expert systems} for various parts of the simulation, such as information collection for observation and casualty estimation. This will enable us to improve the accuracy and realism of the simulation results, while LLM remains solely responsible for decision-making.

Furthermore, we are interested in developing \textbf{more realistic simulations of individual soldiers} beyond just adopting prompting with personalized information. This will enable us to capture the personal experiences of soldiers on the battlefield more accurately and comprehensively.

Finally, we plan to \textbf{explore the interaction between commanding agents and soldier agents}, enabling soldier agents to not only follow commands but also actively affect the decision-making process. This will provide insights into the dynamics of command and control in historical battles and enhance the realism of the simulation.

In summary, our future work aims to extend and enhance our approach to provide even more realistic and comprehensive simulations of historical battles, capturing the complexities and dynamics of warfare and providing valuable insights into the strategies and tactics employed by both armies.

\bibliographystyle{unsrt}  
\bibliography{references} 

\begin{thebibliography}{10}

\bibitem{xi2023rise}
Zhiheng Xi, Wenxiang Chen, Xin Guo, Wei He, Yiwen Ding, Boyang Hong, Ming Zhang, Junzhe Wang, Senjie Jin, Enyu Zhou, et~al.
\newblock The rise and potential of large language model based agents: A survey.
\newblock {\em arXiv preprint arXiv:2309.07864}, 2023.

\bibitem{zhao2023survey}
Wayne~Xin Zhao, Kun Zhou, Junyi Li, Tianyi Tang, Xiaolei Wang, Yupeng Hou, Yingqian Min, Beichen Zhang, Junjie Zhang, Zican Dong, et~al.
\newblock A survey of large language models.
\newblock {\em arXiv preprint arXiv:2303.18223}, 2023.

\bibitem{fan2023bibliometric}
Lizhou Fan, Lingyao Li, Zihui Ma, Sanggyu Lee, Huizi Yu, and Libby Hemphill.
\newblock A bibliometric review of large language models research from 2017 to 2023.
\newblock {\em arXiv preprint arXiv:2304.02020}, 2023.

\bibitem{huang2022towards}
Jie Huang and Kevin Chen-Chuan Chang.
\newblock Towards reasoning in large language models: A survey.
\newblock {\em arXiv preprint arXiv:2212.10403}, 2022.

\bibitem{jin2024impact}
Mingyu Jin, Qinkai Yu, Haiyan Zhao, Wenyue Hua, Yanda Meng, Yongfeng Zhang, Mengnan Du, et~al.
\newblock The impact of reasoning step length on large language models.
\newblock {\em arXiv preprint arXiv:2401.04925}, 2024.

\bibitem{zhou2023instruction}
Jeffrey Zhou, Tianjian Lu, Swaroop Mishra, Siddhartha Brahma, Sujoy Basu, Yi~Luan, Denny Zhou, and Le~Hou.
\newblock Instruction-following evaluation for large language models.
\newblock {\em arXiv preprint arXiv:2311.07911}, 2023.

\bibitem{zeng2023evaluating}
Zhiyuan Zeng, Jiatong Yu, Tianyu Gao, Yu~Meng, Tanya Goyal, and Danqi Chen.
\newblock Evaluating large language models at evaluating instruction following.
\newblock {\em arXiv preprint arXiv:2310.07641}, 2023.

\bibitem{wang2024llms}
Boshi Wang, Hao Fang, Jason Eisner, Benjamin Van~Durme, and Yu~Su.
\newblock Llms in the imaginarium: Tool learning through simulated trial and error.
\newblock {\em arXiv preprint arXiv:2403.04746}, 2024.

\bibitem{wang2024tool}
Chenyu Wang, Weixin Luo, Qianyu Chen, Haonan Mai, Jindi Guo, Sixun Dong, Zhengxin Li, Lin Ma, Shenghua Gao, et~al.
\newblock Tool-lmm: A large multi-modal model for tool agent learning.
\newblock {\em arXiv preprint arXiv:2401.10727}, 2024.

\bibitem{shen2024small}
Weizhou Shen, Chenliang Li, Hongzhan Chen, Ming Yan, Xiaojun Quan, Hehong Chen, Ji~Zhang, and Fei Huang.
\newblock Small llms are weak tool learners: A multi-llm agent.
\newblock {\em arXiv preprint arXiv:2401.07324}, 2024.

\bibitem{zhang2024vision}
Jingyi Zhang, Jiaxing Huang, Sheng Jin, and Shijian Lu.
\newblock Vision-language models for vision tasks: A survey.
\newblock {\em IEEE Transactions on Pattern Analysis and Machine Intelligence}, 2024.

\bibitem{durante2024agent}
Zane Durante, Qiuyuan Huang, Naoki Wake, Ran Gong, Jae~Sung Park, Bidipta Sarkar, Rohan Taori, Yusuke Noda, Demetri Terzopoulos, Yejin Choi, et~al.
\newblock Agent ai: Surveying the horizons of multimodal interaction.
\newblock {\em arXiv preprint arXiv:2401.03568}, 2024.

\bibitem{xie2024large}
Junlin Xie, Zhihong Chen, Ruifei Zhang, Xiang Wan, and Guanbin Li.
\newblock Large multimodal agents: A survey.
\newblock {\em arXiv preprint arXiv:2402.15116}, 2024.

\bibitem{du2023improving}
Yilun Du, Shuang Li, Antonio Torralba, Joshua~B. Tenenbaum, and Igor Mordatch.
\newblock Improving factuality and reasoning in language models through multiagent debate, 2023.

\bibitem{chan2023chateval}
Chi-Min Chan, Weize Chen, Yusheng Su, Jianxuan Yu, Wei Xue, Shanghang Zhang, Jie Fu, and Zhiyuan Liu.
\newblock Chateval: Towards better llm-based evaluators through multi-agent debate, 2023.

\bibitem{sun2023corex}
Qiushi Sun, Zhangyue Yin, Xiang Li, Zhiyong Wu, Xipeng Qiu, and Lingpeng Kong.
\newblock Corex: Pushing the boundaries of complex reasoning through multi-model collaboration, 2023.

\bibitem{liang2023encouraging}
Tian Liang, Zhiwei He, Wenxiang Jiao, Xing Wang, Yan Wang, Rui Wang, Yujiu Yang, Zhaopeng Tu, and Shuming Shi.
\newblock Encouraging divergent thinking in large language models through multi-agent debate, 2023.

\bibitem{hong2023metagpt}
Sirui Hong, Xiawu Zheng, Jonathan Chen, Yuheng Cheng, Ceyao Zhang, Zili Wang, Steven Ka~Shing Yau, Zijuan Lin, Liyang Zhou, Chenyu Ran, et~al.
\newblock Metagpt: Meta programming for multi-agent collaborative framework.
\newblock {\em arXiv preprint arXiv:2308.00352}, 2023.

\bibitem{liu2023bolaa}
Zhiwei Liu, Weiran Yao, Jianguo Zhang, Le~Xue, Shelby Heinecke, Rithesh Murthy, Yihao Feng, Zeyuan Chen, Juan~Carlos Niebles, Devansh Arpit, et~al.
\newblock Bolaa: Benchmarking and orchestrating llm-augmented autonomous agents.
\newblock {\em arXiv preprint arXiv:2308.05960}, 2023.

\bibitem{ge2023openagi}
Yingqiang Ge, Wenyue Hua, Kai Mei, Jianchao Ji, Juntao Tan, Shuyuan Xu, Zelong Li, and Yongfeng Zhang.
\newblock Open{AGI}: When {LLM} meets domain experts.
\newblock In {\em Thirty-seventh Conference on Neural Information Processing Systems}, 2023.

\bibitem{yang2023appagent}
Zhao Yang, Jiaxuan Liu, Yucheng Han, Xin Chen, Zebiao Huang, Bin Fu, and Gang Yu.
\newblock Appagent: Multimodal agents as smartphone users.
\newblock {\em arXiv preprint arXiv:2312.13771}, 2023.

\bibitem{mei2024llm}
Kai Mei, Zelong Li, Shuyuan Xu, Ruosong Ye, Yingqiang Ge, and Yongfeng Zhang.
\newblock Llm agent operating system.
\newblock {\em arXiv preprint arXiv:2403.16971}, 2024.

\bibitem{ge2023llm}
Yingqiang Ge, Yujie Ren, Wenyue Hua, Shuyuan Xu, Juntao Tan, and Yongfeng Zhang.
\newblock Llm as os, agents as apps: Envisioning aios, agents and the aios-agent ecosystem.
\newblock {\em arXiv e-prints}, pages arXiv--2312, 2023.

\bibitem{gong2023mindagent}
Ran Gong, Qiuyuan Huang, Xiaojian Ma, Hoi Vo, Zane Durante, Yusuke Noda, Zilong Zheng, Song-Chun Zhu, Demetri Terzopoulos, Li~Fei-Fei, et~al.
\newblock Mindagent: Emergent gaming interaction.
\newblock {\em arXiv preprint arXiv:2309.09971}, 2023.

\bibitem{xu2023exploring}
Yuzhuang Xu, Shuo Wang, Peng Li, Fuwen Luo, Xiaolong Wang, Weidong Liu, and Yang Liu.
\newblock Exploring large language models for communication games: An empirical study on werewolf.
\newblock {\em arXiv preprint arXiv:2309.04658}, 2023.

\bibitem{lan2023llm}
Yihuai Lan, Zhiqiang Hu, Lei Wang, Yang Wang, Deheng Ye, Peilin Zhao, Ee-Peng Lim, Hui Xiong, and Hao Wang.
\newblock Llm-based agent society investigation: Collaboration and confrontation in avalon gameplay.
\newblock {\em arXiv preprint arXiv:2310.14985}, 2023.

\bibitem{hu2024survey}
Sihao Hu, Tiansheng Huang, Fatih Ilhan, Selim Tekin, Gaowen Liu, Ramana Kompella, and Ling Liu.
\newblock A survey on large language model-based game agents.
\newblock {\em arXiv preprint arXiv:2404.02039}, 2024.

\bibitem{pang2024self}
Xianghe Pang, Shuo Tang, Rui Ye, Yuxin Xiong, Bolun Zhang, Yanfeng Wang, and Siheng Chen.
\newblock Self-alignment of large language models via multi-agent social simulation.
\newblock In {\em ICLR 2024 Workshop on Large Language Model (LLM) Agents}.

\bibitem{zhou2024real}
Xuhui Zhou, Zhe Su, Tiwalayo Eisape, Hyunwoo Kim, and Maarten Sap.
\newblock Is this the real life? is this just fantasy? the misleading success of simulating social interactions with llms.
\newblock {\em arXiv preprint arXiv:2403.05020}, 2024.

\bibitem{sreedhar2024simulating}
Karthik Sreedhar and Lydia Chilton.
\newblock Simulating human strategic behavior: Comparing single and multi-agent llms.
\newblock {\em arXiv preprint arXiv:2402.08189}, 2024.

\bibitem{xie2024can}
Chengxing Xie, Canyu Chen, Feiran Jia, Ziyu Ye, Kai Shu, Adel Bibi, Ziniu Hu, Philip Torr, Bernard Ghanem, and Guohao Li.
\newblock Can large language model agents simulate human trust behaviors?
\newblock {\em arXiv preprint arXiv:2402.04559}, 2024.

\bibitem{hua2023war}
Wenyue Hua, Lizhou Fan, Lingyao Li, Kai Mei, Jianchao Ji, Yingqiang Ge, Libby Hemphill, and Yongfeng Zhang.
\newblock War and peace (waragent): Large language model-based multi-agent simulation of world wars.
\newblock {\em arXiv preprint arXiv:2311.17227}, 2023.

\bibitem{shopes2011oral}
Linda Shopes.
\newblock Oral history.
\newblock {\em The SAGE handbook of qualitative research}, pages 451--465, 2011.

\bibitem{portelli2002makes}
Alessandro Portelli.
\newblock What makes oral history different.
\newblock In {\em The oral history reader}, pages 77--88. Routledge, 2002.

\bibitem{fan2023nphardeval}
Lizhou Fan, Wenyue Hua, Lingyao Li, Haoyang Ling, Yongfeng Zhang, and Libby Hemphill.
\newblock Nphardeval: Dynamic benchmark on reasoning ability of large language models via complexity classes.
\newblock {\em arXiv preprint arXiv:2312.14890}, 2023.

\bibitem{fan2024nphardeval4v}
Lizhou Fan, Wenyue Hua, Xiang Li, Kaijie Zhu, Mingyu Jin, Lingyao Li, Haoyang Ling, Jinkui Chi, Jindong Wang, Xin Ma, et~al.
\newblock Nphardeval4v: A dynamic reasoning benchmark of multimodal large language models.
\newblock {\em arXiv preprint arXiv:2403.01777}, 2024.

\bibitem{zhang2024llm}
Yadong Zhang, Shaoguang Mao, Tao Ge, Xun Wang, Adrian de~Wynter, Yan Xia, Wenshan Wu, Ting Song, Man Lan, and Furu Wei.
\newblock Llm as a mastermind: A survey of strategic reasoning with large language models.
\newblock {\em arXiv preprint arXiv:2404.01230}, 2024.

\bibitem{park2023generative}
Joon~Sung Park, Joseph~C O'Brien, Carrie~J Cai, Meredith~Ringel Morris, Percy Liang, and Michael~S Bernstein.
\newblock Generative agents: Interactive simulacra of human behavior.
\newblock {\em arXiv preprint arXiv:2304.03442}, 2023.

\bibitem{fu2023improving}
Yao Fu, Hao Peng, Tushar Khot, and Mirella Lapata.
\newblock Improving language model negotiation with self-play and in-context learning from ai feedback.
\newblock {\em arXiv preprint arXiv:2305.10142}, 2023.

\bibitem{bran2023augmenting}
Andres~M Bran, Sam Cox, Oliver Schilter, Carlo Baldassari, Andrew White, and Philippe Schwaller.
\newblock Augmenting large language models with chemistry tools.
\newblock In {\em NeurIPS 2023 AI for Science Workshop}, 2023.

\bibitem{hua2024trustagent}
Wenyue Hua, Xianjun Yang, Zelong Li, Cheng Wei, and Yongfeng Zhang.
\newblock Trustagent: Towards safe and trustworthy llm-based agents through agent constitution.
\newblock {\em arXiv preprint arXiv:2402.01586}, 2024.

\bibitem{chen2023towards}
Kexin Chen, Junyou Li, Kunyi Wang, Yuyang Du, Jiahui Yu, Jiamin Lu, Guangyong Chen, Lanqing Li, Jiezhong Qiu, Qun Fang, et~al.
\newblock Towards an automatic ai agent for reaction condition recommendation in chemical synthesis.
\newblock {\em arXiv preprint arXiv:2311.10776}, 2023.

\bibitem{AppAgent}
Zhao Yang, Jiaxuan Liu, Yucheng Han, Xin Chen, Zebiao Huang, Bin Fu, and Gang Yu.
\newblock Appagent: Multimodal agents as smartphone users.
\newblock {\em arXiv preprint arXiv:2312.13771}, 2023.

\bibitem{AWS}
Amazon~Web Service.
\newblock Generative ai and multi-modal agents in aws: The key to unlocking new value in financial markets, 1 2024.

\bibitem{LLaVAPlus}
Shilong Liu, Hao Cheng, Haotian Liu, Hao Zhang, Feng Li, Tianhe Ren, Xueyan Zou, Jianwei Yang, Hang Su, Jun Zhu, et~al.
\newblock Llava-plus: Learning to use tools for creating multimodal agents.
\newblock {\em arXiv preprint arXiv:2311.05437}, 2023.

\bibitem{CoA}
Zhuosheng Zhan and Aston Zhang.
\newblock You only look at screens: Multimodal chain-of-action agents.
\newblock {\em arXiv preprint arXiv:2309.11436}, 2023.

\bibitem{dickson2002road}
Ted Dickson.
\newblock The road to united states involvement in world war i: A simulation.
\newblock {\em OAH Magazine of History}, 17(1):48--56, 2002.

\bibitem{guetzkow1963simulation}
Harold~Steere Guetzkow, Chadwick~F Alger, and Richard~A Brody.
\newblock Simulation in international relations: Developments for research and teaching.
\newblock {\em (No Title)}, 1963.

\bibitem{hermann1967attempt}
Charles~F Hermann and Margaret~G Hermann.
\newblock An attempt to simulate the outbreak of world war i.
\newblock {\em American Political Science Review}, 61(2):400--416, 1967.

\bibitem{tollefson2008onesaf}
Eric Tollefson, M~Martin, Andrew Fletcher, and ARMY TRADOC ANALYSIS CENTER~MONTEREY CA.
\newblock Onesaf objective system (oos) behavior model verification.
\newblock {\em US Army TRADOC Analysis Center--Monterey, Monterey, CA}, 2008.

\bibitem{hill2004using}
Raymond~R Hill, Lance~E Champagne, and Joseph~C Price.
\newblock Using agent-based simulation and game theory to examine the wwii bay of biscay u-boat campaign.
\newblock {\em The Journal of Defense Modeling and Simulation}, 1(2):99--109, 2004.

\bibitem{ghaffarzadegan2023generative}
Navid Ghaffarzadegan, Aritra Majumdar, Ross Williams, and Niyousha Hosseinichimeh.
\newblock Generative agent-based modeling: Unveiling social system dynamics through coupling mechanistic models with generative artificial intelligence.
\newblock {\em arXiv preprint arXiv:2309.11456}, 2023.

\bibitem{e21pstein1999agent}
Joshua~M Epstein.
\newblock Agent-based computational models and generative social science.
\newblock {\em Complexity}, 1999.

\bibitem{fan2020stigmatization}
Lizhou Fan, Huizi Yu, and Zhanyuan Yin.
\newblock Stigmatization in social media: Documenting and analyzing hate speech for covid-19 on twitter.
\newblock {\em Proceedings of the Association for Information Science and Technology}, 57(1):e313, 2020.

\bibitem{yin2020using}
Zhanyuan Yin, Lizhou Fan, Huizi Yu, and Anne~J Gilliland.
\newblock Using a three-step social media similarity (tsms) mapping method to analyze controversial speech relating to covid-19 in twitter collections.
\newblock In {\em 2020 IEEE International Conference on Big Data (Big Data)}, pages 1949--1953. IEEE, 2020.

\bibitem{li2023chatgpt}
Lingyao Li, Zihui Ma, Lizhou Fan, Sanggyu Lee, Huizi Yu, and Libby Hemphill.
\newblock Chatgpt in education: A discourse analysis of worries and concerns on social media.
\newblock {\em arXiv preprint arXiv:2305.02201}, 2023.

\bibitem{burne2016crecy}
Alfred~H Burne.
\newblock {\em The Crecy War: A Military History of the Hundred Years War from 1337 to the Peace of Bretigny in 1360}.
\newblock Casemate Publishers, 2016.

\bibitem{curry2000battle}
Anne Curry.
\newblock {\em The battle of Agincourt: sources and interpretations}.
\newblock Boydell Press, 2000.

\bibitem{given2001edward}
Chris Given-Wilson and Fran{\c{c}}oise B{\'e}riac.
\newblock Edward iii's prisoners of war: the battle of poitiers and its context.
\newblock {\em The English Historical Review}, 116(468):802--833, 2001.

\bibitem{kiffer2019battle}
Andr{\'e}~Geraque Kiffer.
\newblock {\em Battle Of Falkirk, July 22, 1298}.
\newblock Clube de Autores, 2019.

\bibitem{matsuoka2001computer}
Toshifumi Matsuoka, Takahiro Hasegawa, Yasuhiro Yamada, Tetsuya Tamagawa, and Yuzuru Ashida.
\newblock Computer simulation for sandbox experiments.
\newblock In {\em SEG International Exposition and Annual Meeting}, pages SEG--2001. SEG, 2001.

\bibitem{al2011effects}
Ahmed~A Al~Rowaei, Arnold~H Buss, and Stephen Lieberman.
\newblock The effects of time advance mechanism on simple agent behaviors in combat simulations.
\newblock In {\em Proceedings of the 2011 Winter Simulation Conference (WSC)}, pages 2426--2437. IEEE, 2011.

\bibitem{liu2023dynamic}
Zijun Liu, Yanzhe Zhang, Peng Li, Yang Liu, and Diyi Yang.
\newblock Dynamic llm-agent network: An llm-agent collaboration framework with agent team optimization.
\newblock {\em arXiv preprint arXiv:2310.02170}, 2023.

\bibitem{han2024llm}
Shanshan Han, Qifan Zhang, Yuhang Yao, Weizhao Jin, Zhaozhuo Xu, and Chaoyang He.
\newblock Llm multi-agent systems: Challenges and open problems.
\newblock {\em arXiv preprint arXiv:2402.03578}, 2024.

\bibitem{claude3}
Anthropic.
\newblock The claude 3 model family: Opus, sonnet, haiku.
\newblock \url{https://www-cdn.anthropic.com/de8ba9b01c9ab7cbabf5c33b80b7bbc618857627/Model_Card_Claude_3.pdf}, 2024.

\bibitem{achiam2023gpt}
Josh Achiam, Steven Adler, Sandhini Agarwal, Lama Ahmad, Ilge Akkaya, Florencia~Leoni Aleman, Diogo Almeida, Janko Altenschmidt, Sam Altman, Shyamal Anadkat, et~al.
\newblock Gpt-4 technical report.
\newblock {\em arXiv preprint arXiv:2303.08774}, 2023.

\bibitem{gpt4v}
OpenAI.
\newblock Gpt-4v(ision) system card.
\newblock \url{https://cdn.openai.com/papers/GPTV_System_Card.pdf}, 2023.

\end{thebibliography}
\end{document}